\documentstyle[12pt]{article}
\def\zid{1\kern-0.36em\llap~1}

\newcommand{\beq}{\begin{equation}}
\newcommand{\ber}{\begin{eqnarray}}
\newcommand{\eeq}{\end{equation}}
\newcommand{\eer}{\end{eqnarray}}

\begin{document}

\begin{titlepage}
\rightline{[SUNY BING 12/22/00 I] }  \rightline{ hep-ph/0101303}
\vspace{2mm}
\begin{center}
{\bf \hspace{1.85 cm} USE OF $\Lambda_b$ POLARIMETRY IN TOP QUARK
\newline SPIN-CORRELATION FUNCTIONS }\\ \vspace{2mm} Charles A.
Nelson\footnote{Electronic address: cnelson @ binghamton.edu  } \\
{\it Department of Physics, State University of New York at
Binghamton\\ Binghamton, N.Y. 13902-6016}\\[2mm]
\end{center}


\begin{abstract}

Due to the absence of hadronization effects and the large $m_t$
mass, top quark decay will be uniquely sensitive to fundamental
electroweak physics at the Tevatron, at the LHC, and at a future
linear collider. A ``complete measurement" of the four helicity
amplitudes in $t \rightarrow W^+ b$ decay is possible by the
combined use of $\Lambda_b$ and $W$ polarimetry in stage-two
spin-correlation functions (S2SC).  In this paper, the most
general Lorentz-invariant decay density matrix is obtained for the
decay sequence $t\rightarrow W^{+}b$ where $b\rightarrow
l^{-}\bar{\nu}c$ and $W^{+}\rightarrow l^{+}\nu _{l}$ [ or
$W^{+}\rightarrow j\overline{_{d}}j_{u}$ ], and likewise for
$\bar{t} \rightarrow W^- \bar{b}$.  These density matrices are
expressed in terms of $b$-polarimetry helicity parameters which
enable a unique determination of the relative phases among the
$A(\lambda_{W^+},\lambda_b)$ amplitudes. Thereby, S2SC
distributions and single-sided $b$-$W$-interference distributions
are expressed in terms of these parameters. The four
$b$-polarimetry helicity parameters involving the $A(-1,-1/2)$
amplitude are considered in detail.  $\Lambda_b$ polarimetry
signatures will not be suppressed in top quark analyses when final
$\bar{\nu}$ angles-and-energy variables are used for $b\rightarrow
l^{-}\bar{\nu}c$.

\end{abstract}

\end{titlepage}

\section{ Introduction:}

Assuming only $W$-polarimetry information\cite{fnal}, in
Ref.\cite{nklm1} we used the helicity formalism to derive
stage-two spin-correlation functions for top quark decays.
However, a complete determination of the decay amplitudes requires
information from $b$-polarimetry, such as from $\Lambda_b$ decays.
In this paper we accordingly generalize the earlier results to
enable measurement of the relative phase of $b_L$ and $b_R$
amplitudes by $\Lambda_b$ polarimetry [3-6][7-10].

The significant point \cite{nklm2} is that by this technique a
``complete measurement" of the four helicity amplitudes $A(
\lambda_{W^+}, \lambda_b )$ in $t \rightarrow W^+ b$ decay is
possible: If only $b_L$ coupling's existed, there would be only 2
amplitudes, so 3 quantities would determine $t \rightarrow W^+ b
$: measurement of the magnitude's ratio $r_a^L \equiv \frac{|A(-
1,-\frac 12)|}{|A(0,-\frac 12)|}$, of the L-handed relative phase
$ \beta _a^L$, and of the partial width $\Gamma$.  But, since $m_b
\neq 0$, there are 2 more amplitudes, so to achieve an ``almost''
complete measurement \cite{nklm1}, 3 additional quantities must be
determined:  e.g. $r_a^R \equiv \frac{|A(1,\frac 12)|}{|A(0,\frac
12)|}$, the R-handed relative phase $ \beta _a^R$, and a L-R
magnitude's ratio $ r_a^1 \equiv \frac{|A(1,\frac
12)|}{|A(0,-\frac 12)|}$. However, a further measurement is
required to determine the relative phase of the $b_L$ and $b_R$
amplitudes, such as $\gamma_{+}^a\equiv \phi _1^{a}-\phi _o^a$
which is between the two $\lambda_b = \pm 1/2$ amplitudes with the
largest moduli in the standard model (SM). See the upper sketch in
Fig. 1 for definition of these relative phases \cite{cohen}. By a
combined use of $\Lambda_b$ and $W^+$ polarimetry, it is possible
to obtain this missing phase through a measurement of the
interference between the $b_L$ and $b_R$ amplitudes in $t
\rightarrow  W^+ b $ with $b \rightarrow c l^- \bar \nu$ where the
$b$-quark is required to occur in a bound state in the $\Lambda_b$
mass region. For instance, the helicity parameters
\begin{equation}
{\epsilon _{+}}\equiv \frac{1}{{\Gamma }}\left| A(1,\frac{%
1}{2})\right| \left| A(0,-\frac{1}{2})\right| \cos {\gamma _{+}}%
,\qquad {\epsilon _{+}}^{^{\prime }}\equiv \frac{1}{{%
\Gamma }}\left| A(1,\frac{1}{2})\right| \left|
A(0,-\frac{1}{2})\right| \sin {\gamma _{+}}
\end{equation}
appear in the stage-two spin-correlation distributions (S2SC),
such as (56,59) below, and in single-sided sequential-decay
distributions, such as (62).  Primed helicity parameters depend on
``sine" functions of the relative phases and so can be used to
test for $\tilde{T}_{FS}$-violation \cite{nklm1,adler}.

At present, the available experimental and theoretical information
regarding future application of $\Lambda_b$ polarimetry is
promising: The pre-measurement, heavy-quark-effective-theory
predictions (HQET) [14-16] for $Z^o$ decays were $<P_{\Lambda_b}
> \approx -0.7 \pm 0.1$ with small QCD corrections.  Initial
LEP1 measurements in 1996-7 were significantly smaller but with
large errors [ALEPH reported $(-0.23\pm 0.25)$ \cite{aleph} and
DELPHI reported $(- 0.08\pm 0.39)$ \cite{delphi} ]. The later OPAL
result \cite{opal} of $ <P_{\Lambda_b} > = -0.56(+0.20/-0.13)\pm
0.09$ and the more recent DELPHI result \cite{delphi2} of  $
<P_{\Lambda_b} > = -0.49(+0.32/-0.30)\pm 0.17$ are both consistent
with a large polarization as in the HQET. The OPAL measurement of
the product of branching ratios is $B(b \rightarrow \Lambda_b X)
B(\Lambda_b \rightarrow \Lambda X ) = 2.67 \pm 0.38 (+0.67/-0.60)
\%$. There is no information on $\Lambda_b$ decays from CESR or
from the on-going B-factory $CP$-violation experiments because of
their choice of operating at an upsilon resonance, the $\Upsilon
(4S)$, at too low an $E_{cm}$ energy so as to maximize B-meson
production.

Throughout this S2SC analysis, we use the following notations: Per
the $A(\lambda_{W^+},\lambda_b)$ amplitudes, lowercase ``a" or
``1" subscripts denote quantities describing the {\bf $t$-decay
side}, whereas per the use of $B(\lambda_{W^-},\lambda_{\bar{b}})$
amplitudes for the $CP$-conjugate $\bar{t} \rightarrow W^-
\bar{b}$ decay, lowercase ``b" or ``2" subscripts similarly denote
quantities for the {\bf $\bar{t}$-decay side}. To be clear versus
the notation for the fundamental $t \rightarrow W^+ b$ process, we
will use ``barred" accents for helicity parameters and other
quantities describing the $CP$-conjugate sequential-decay $
\bar{t} \rightarrow W^- \bar{b} \rightarrow (l^-
\bar{\nu_{l}})(l^+ \nu \bar{c}) $, see Fig. 2 and (19-32). For the
second-stage of $ t \rightarrow W^+ b \rightarrow (l^+
\nu_{l})(l^- \bar{\nu} c)$, we use {\bf tilde accents} to denote
angles for $W^+ \rightarrow l^+ \nu_{l}$ [ or $W^{+}\rightarrow
j\overline{_{d}}j_{u}$ ] and for density matrices {\bf depending
only} on $W$-polarimetry.  We use {\bf hat accents} to denote
angles for $b \rightarrow l^- \bar{\nu} c$ and for density
matrices {\bf depending} on $b$-polarimetry. For the second-stage
of the $CP$-conjugate sequence we correspondingly use these
``tilde" and ``hat" accents. Invariance under any fundamental
discrete symmetry such as $CP$, $T$, or $CPT$ is not assumed in
this analysis nor in the earlier papers because the framework is
the helicity formalism. A brief introduction to this accent
labeling can be obtained by inspection of the figures.  We also
use a {\bf slash notation} for the ``full" or
double-sequential-decay density matrices which use both $b$-quark
and $W$-polarimetry, e.g. (30).

Below, in Sec. 2, we construct the most general Lorentz-invariant
decay density matrix for the decay sequence $t\rightarrow W^{+}b$
where $b\rightarrow l^{-}\bar{\nu}c$ and also $W^{+}\rightarrow
l^{+}\nu _{l}$ \ [ or $W^{+}\rightarrow j\overline{_{d}}j_{u}$ ].
The corresponding quantities are also obtained for the
$CP$-conjugate sequential-decay.  For it and associated with Fig.
2, we list explicitly the ``barred" formulas for its helicity
parameters and relative phases. Simple $CP$ tests were treated in
\cite{nklm1}.  In Sec. 3, we then generalize the derivation of the
earlier stage-two spin-correlation functions in the case of both
$W$-polarimetry and $b$-polarimetry. We similarly generalize the
single-sided sequential-decay distributions.

In Sec. 4, the $b$-polarimetry helicity parameters $\epsilon_{-},
\kappa_{1}$ involving the L-handed $b$-quark amplitude
$A(-1,-1/2)$ are considered in detail regarding tests for
single-additional Lorentz-invariant couplings. We also consider
the analogous ``primed parameters" regarding signatures for
$\tilde{T}_{FS}$-violation. Those involving $A(0,-1/2)$ were
treated in \cite{cohen,adler}. In the SM and at the ($S+P$) and
($f_M+f_E$) ambiguous moduli points, the values of the four
$b$-polarimetry interference parameters $\epsilon_{\pm},
\kappa_{o,1}$ are small$\sim 1\%$. However at low-effective mass
scales ($ < 320GeV$), the values of these parameters can be large,
$0.1$ to $0.4$ versus their unitarity limit of $0.5$, for
single-additional Lorentz structures having sizable R-handed
$b$-quark amplitudes, c.f. Figs. 9-10 below and Figs. 8-9 in
\cite{cohen}.

In the near future, $\Lambda_b$ polarimetry could be used in top
quark spin-correlation analyses at the Tevatron, at the LHC, and
at a future linear collider \cite{nlc}. If the
heavy-quark-effective theory prediction is correct, then depending
on the dynamics occurring in top quark decay and on good
detector/accelerator polarimetry capabilities, we think that
$\Lambda_b$ polarimetry could be a very important technique in
studying fundamental electroweak physics through top quark decay
processes.

\section{Use of $\Lambda_b$ Polarimetry in Sequential-Decay
\newline Density Matrices}

In order to include $b$-polarimetry, we generalize the derivation
of state-two-spin-correlation functions given in \cite{nklm1}.

In the  $t$ rest frame, the matrix element for $t \rightarrow
W^{+} b$ is
\begin{equation}
\langle \theta _1^t ,\phi _1^t ,\lambda _{W^{+} } ,\lambda _b
|\frac 12,\lambda _1\rangle =D_{\lambda _1,\mu }^{(1/2)*}(\phi
_1^t ,\theta _1^t ,0)A\left( \lambda _{W^{+} } ,\lambda _b \right)
\end{equation}
where $\mu =\lambda _{W^{+} } -\lambda _b $ in terms of the $W^+$
and $b$-quark helicities. The asterisk denotes complex
conjugation.  The final $W^{+}$ momentum is in the $\theta _1^t
,\phi _1^t$ direction and the $b$-quark momentum is in the
opposite direction. $\lambda_1$ gives the $t$-quark's spin
component quantized along the $z_{1}^t$ axis in Fig. 3.  So, upon
a boost back to the $(t \bar{t})$ center-of-mass frame, or on to
the $\bar{t}$ rest frame, $\lambda_1$ also specifies the helicity
of the $t$-quark. For the $CP$-conjugate process, $\bar t
\rightarrow W^{-} \bar b$, in the $\bar t$ rest frame
\begin{equation}
\langle \theta _2^t ,\phi _2^t ,\lambda _{W^{-} },\lambda _{\bar
b}|\frac 12,\lambda _2\rangle =D_{\lambda _2,\bar \mu
}^{(1/2)*}(\phi _2^t ,\theta _2^t ,0)B\left( \lambda _{W^{-}
},\lambda _{\bar b}\right)
\end{equation}
with $\bar \mu =\lambda
_{W^{-}}-\lambda _{\bar b}$ and by the analogous argument
$\lambda_2$ is the $\bar t$ helicity.

To use $\Lambda_b$-polarimetry in the S2SC functions, we consider
the decay sequence $t\rightarrow W^{+}b$ followed by
 $b\rightarrow l^{-}\bar{\nu} X $.  In Figs. 3 and 4, the spherical angles
$\widehat{\theta _1^t}$ and $\widehat{\phi _1^t}$  describe the
$b$ momentum in the ``first stage" $t\rightarrow W^{+}b$.  Note
that $\widehat{\theta _1^t} = \pi - {\theta _1^t}$.  Note that
$\hat{\phi}$ is the important opening-angle between the $t$-quark
and $\bar{t}$-quark decay planes, and $\widehat{\theta _2^t} = \pi
- {\theta _2^t}$.

In Fig. 4, the angles $ \widehat{\theta _a}$ and $\widehat{\phi
_a}$ describe the $l^{-}$ momentum in the ``second stage"
$b\rightarrow l^{-}\bar{\nu} X $ when the boost is from the $t_1$
rest frame. In Fig. 5 the spherical angles $\widehat{\theta _b} $,
$\widehat{\phi _b}$ similarly specify the $l^{+ }$ momentum in the
$\bar{b}$ rest frame when the boost is from the $\bar t_2$ rest
frame.

As shown in Fig. 6, when the boost to these $b$ and $\bar{b}$ rest
frames is directly from the $(t \bar t)_{cm} $ center-of-mass
frame, we use respectively the subscripts ``$1,2$" in place of the
subscripts ``$a,b$".
Physically these angles, $\widehat \theta _a$, $%
\widehat \phi _a$ and $\widehat \theta _1$, $%
\widehat \phi _1$, are simply related by a Wigner-rotation, see
below following (58).  For the $CP$-conjugate mode, one only needs
to change the subscripts $a \rightarrow b, 1 \rightarrow 2$.

For $W$-polarimetry, we proceed as in \cite{nklm1} and use the
angles shown in Figs. 7 and 8.  In Fig. 7, the angles $\theta
_1^t$, $\phi _1^t$ and $ \widetilde{\theta _a}$,$\widetilde{\phi
_a}$ describe the respective stages in the sequential decay
$t\rightarrow W^{+}b$ followed by $W^{+}\rightarrow j_{\bar d}j_u$
[ or $W^{+}\rightarrow l^{+}\nu $]. For the hadronic $W^{+}$ decay
mode, we use the notation that the momentum of the charge
$\frac{1}{3} e$ jet is denoted by $j_{\bar d}$ and the momentum of
the charge $\frac{2}{3} e$ jet by $j_{u}$. Similarly, in Fig. 8,
$\widetilde{\theta _b} $, $\widetilde{\phi _b}$ specify the $j_d$
jet (or the $l^{- }$) momentum in the $W^{-}$rest frame. When the
boost to these $W^{\pm}$ rest frames is directly from the $(t \bar
t)_{cm} $ center-of-mass frame, we also use subscripts ``$1,2$" in
place of ``$a,b$".  This is shown explicitly in Fig. 4 of
\cite{nklm1}; we omit that figure here for it is exactly analogous
to Fig. 6 of the present paper.  The angles in the $W^{+}$ rest
frame, $\tilde \theta _a$, $ \tilde \phi _a$ and $\tilde \theta
_1$, $ \tilde \phi _1$, are simply related by a Wigner-rotation [
see below following (58) ].  For the $CP$-conjugate mode's $W^{-}$
rest frame, one
 again only needs to change the subscripts $a \rightarrow b, 1
\rightarrow 2$.

In the $W^{+}$ rest frame, the matrix element for $W^{+}
\rightarrow l^+ \nu$ [ or for $W^{+} \rightarrow j_{\bar d} j_{u}$
] is \cite{nklm1}
\begin{equation}
\langle \tilde \theta _a ,\tilde \phi _a ,\lambda _{l^+} ,\lambda
_{\nu} | 1,\lambda _{W^+} \rangle =D_{\lambda _{W^+},1 }^{1*}
 (\tilde \phi
_a ,\tilde \theta _a ,0)c
\end{equation}
since $\lambda _{\nu}= - \frac{1}{2}, \lambda _{l^+}=
\frac{1}{2}$, neglecting $(\frac{m_l}{m_W})$ corrections [
neglecting $(\frac{ m_{jet} }{m_W})$ corrections].  Since the
amplitude ``c" in this matrix element is then independent of the
helicities, we will usually suppress it in the following formulas
since it only effects the overall normalization. We will use below
$$ \rho _{\lambda _1\lambda _1^{^{\prime }};\lambda _W\lambda
_W^{^{\prime }}}(t\rightarrow W^{+}b)=\sum_{\lambda _b=\mp
1/2}D_{\lambda _1,\mu }^{(1/2)*}(\phi _1^t,\theta
_1^t,0)D_{\lambda _1^{^{\prime }},\mu ^{^{\prime }}}^{(1/2)}(\phi
_1^t,\theta _1^t,0)A(\lambda _W,\lambda _b)A(\lambda _W^{^{\prime
}},\lambda _b)^{*} $$ $$ \widetilde{ \rho} _{\lambda _W\lambda
_W^{^{\prime }}} (W^{+}\rightarrow l^{+}\nu )=D_{\lambda
_W,1}^{1*}(\widetilde{\phi _a},\widetilde{\theta
_a}%
,0)D_{\lambda _W^{^{\prime }},1}^1(\widetilde{\phi
_a},\widetilde{\theta _a}%
,0) |c|^2 $$
where $\mu =\lambda _{W^{+}}-\lambda _{b}$ and $\mu
^{^{\prime }}=\lambda _{W^{+}}-\lambda _{b}^{^{\prime }}$.

\subsection{ Sequential-decay density matrices}

{\bf Case: Only $b$-quark polarimetry:}

The decay density matrix for the first stage of the decay sequence
when the $W$ helicities are summed over is
\begin{equation}
\hat{\rho}_{\lambda _{1}\lambda _{1}^{^{\prime }},\lambda
_{b}\lambda _{b}^{^{\prime }}}(t\rightarrow W^{+}b)=\sum_{\lambda
_{W^{+}}=\pm 1,0}D_{\lambda _{1}\mu }^{1/2\ast }(\phi
_{1}^{t},\theta _{1}^{t},0)D_{\lambda _{1}^{^{\prime }}\mu
^{^{\prime }}}^{1/2}(\phi _{1}^{t},\theta _{1}^{t},0)A(\lambda
_{W^{+}},\lambda _{b})A^{\ast }(\lambda _{W^{+}},\lambda
_{b}^{^{\prime }})
\end{equation}
where $\mu =\lambda _{W^{+}}-\lambda _{b}$ and $\mu ^{^{\prime
}}=\lambda _{W^{+}}-\lambda _{b}^{^{\prime }}$.

Similarly for the second stage of the decay sequence, the decay
density matrix is [ c.f. Eqs.(3-9) in \cite{jd1} ]
\begin{eqnarray}
\hat{\rho}_{\lambda _{b}\lambda _{b}^{^{\prime }}}(b &\rightarrow & l^{-}%
\bar{\nu}c)=\hat{\rho}_{\lambda _{b}\lambda _{b}^{^{\prime
}}}(\widehat{\theta _{a}},\widehat{\phi _{a}},E_{l}) \nonumber \\
&=&\sum_{\Lambda =\pm \frac{1}{2}}D_{\lambda _{b}\Lambda }^{1/2\ast }(%
\widehat{\phi _{a}},\widehat{\theta _{a}},0)D_{\lambda
_{b}^{^{\prime }}\Lambda }^{1/2}(\widehat{\phi
_{a}},\widehat{\theta _{a}},0)\left| R_{\Lambda
}^{b}(E_{l})\right| ^{2}
\end{eqnarray}

As discussed in the Appendix, the two $\left| R_{\Lambda
}^{b}(E_{l})\right|^2$ factors can be expressed in terms of
formulas for the final lepton energy spectra \cite{jd1} in the $
b\rightarrow l^{-}\overline{\nu }c$ decay in the $\Lambda_b$ mass
region. Different methods to determine the $\Lambda_b$
polarization have been investigated and used for $\Lambda_b$'s
arising from $Z^o$ decays. These include spectra measures such as
$<E_l>/<E_{\bar{\nu}}>$ and $<E_l^n>/<E_{\bar{\nu}}^n>$
[3-6,8-10], and spin-momentum correlation measures
\cite{koerner2}.

When the final $\bar{\nu}$ angles-and-energy-distribution are
used, the $\Lambda_b$ polarimetry signatures in all the elements
of the composite decay density matrix $\widehat{\mathbf{R}}$ of
(13), and similarly for $\widehat{\mathbf{\not\!{R}}}$ of (32),
will not be suppressed by the ratio $\frac{ \left| R_{+}\right|
^{2}-\left| R_{-}\right| ^{2} } { \left| R_{+}\right| ^{2}+\left|
R_{-}\right| ^{2}}$ since the $\left| R_{-}\right| ^{2}$ term then
vanishes, see the Appendix. Thereby, $\left| R_{+}\right| ^{2}$
can be completely factored out of $\widehat{\mathbf{R}}$ and
$\widehat{\mathbf{\not\!{R}}}$. For the $CP$-conjugate mode when
the final $\nu$ angles-and-energy-distribution are used, the
analogous suppression factor is absent in all the elements of
$\overline{\widehat{\mathbf{R}}}$ of (14), and of
$\overline{\widehat{\mathbf{\not\!{R}}}}$ of (45), and $\left|
\overline{R_{-}}\right| ^{2}$ completely factors out of
$\overline{\widehat{\mathbf{R}}}$ and
$\overline{\widehat{\mathbf{\not\!{R}}}}$.  This same
factorization occurs in the S2SC distributions, e.g. (60,61), and
in the single-sided $b$-$W$-interference distributions, e.g.
(63-67).

Using (4) and (6), we define the composite decay density matrix
for $t\rightarrow W^{+}b\rightarrow W^{+}(l^{-}\bar{\nu}c)$ by
\begin{equation}
\widehat{R}_{\lambda _{1}\lambda _{1}^{^{\prime }}}(\theta
_{1}^{t},\phi _{1}^{t};\widehat{\theta _{a}},\widehat{\phi
_{a}})=\sum_{\lambda _{b}\lambda _{b}^{^{\prime
}}}\hat{\rho}_{\lambda _{1}\lambda _{1}^{^{\prime
}},\lambda _{b}\lambda _{b}^{^{\prime }}}(t\rightarrow W^{+}b)\hat{\rho}%
_{\lambda _{b}\lambda _{b}^{^{\prime }}}(b\rightarrow
l^{-}\bar{\nu}c)
\end{equation}
where the summation is understood to be over $\lambda _{b}=\pm \frac{1}{2}%
,\lambda _{b}^{^{\prime }}=\pm \frac{1}{2}$. \ This gives
\begin{equation}
\begin{array}{c}
\widehat{R}_{\lambda _{1}\lambda _{1}^{^{\prime }}}=e^{i(\lambda
_{1}-\lambda _{1}^{^{\prime }})\phi _{1}^{t}}\sum_{\lambda
_{W^{+}}=\pm 1,0} \sum_{\lambda _{b}\lambda _{b}^{^{\prime }}}
\{d_{\lambda _{1},\lambda _{W^{+}}-\lambda
_{b}}^{\frac{1}{2}}(\theta _{1}^{t})d_{\lambda
_{1}^{^{\prime }},\lambda _{W^{+}}-\lambda _{b}^{^{\prime }}}^{\frac{1}{2}%
}(\theta _{1}^{t})A(\lambda _{W^{+}},\lambda _{b})A^{\ast
}(\lambda _{W^{+}},\lambda _{b}^{^{\prime }}) \\ e^{i(\lambda
_{b}-\lambda _{b}^{^{\prime }})\widehat{\phi _{a}}}\left[
\sum_{\Lambda =\pm \frac{1}{2}}d_{\lambda _{b},\Lambda }^{\frac{1}{2}}(%
\widehat{\theta _{a}})d_{\lambda _{b}^{^{\prime }},\Lambda }^{\frac{1}{2}}(%
\widehat{\theta _{a}})\left| R_{\Lambda }^{b}(E_{l})\right|
^{2}\right] \}
\end{array}
\end{equation}
In later equations, we will often use the simplifying notation
that
\begin{equation}
\left| R_{\pm }\right| ^{2}\equiv \left| R_{\pm \frac{1}{2}%
}^{b}(E_{l})\right| ^{2}
\end{equation}

The two diagonal elements of the matrix $ \widehat{R}_{\lambda
_{1}\lambda _{1}^{^{\prime }}} $, with $\lambda _{1},\lambda
_{1}^{^{\prime }}=\pm \frac{1}{2},\pm \frac{1}{2}$ , are purely
real.  They can be written in terms of the angles in Fig. 4 and in
terms of the helicity parameters defined in
\cite{nklm1,cohen,adler}.
\begin{equation}
\begin{array}{c}
\widehat{\mathbf{R}}_{\pm \pm }=\frac{\Gamma }{4}\{\left[ \left|
R_{+}\right| ^{2}+\left| R_{-}\right| ^{2}\right] \left( 1\pm
\zeta \cos \theta _{1}^{t}\right) -\cos \widehat{\theta
_{a}}\left[ \left| R_{+}\right| ^{2}-\left| R_{-}\right|
^{2}\right] \left( \xi \pm \sigma \cos \theta _{1}^{t}\right)  \\
\mp 2\sin \widehat{\theta _{a}}\sin \theta _{1}^{t}\left[ \left|
R_{+}\right| ^{2}-\left| R_{-}\right| ^{2}\right] \left( \kappa
_{o}\cos \widehat{\phi _{a}}-\kappa _{o}^{^{\prime }}\sin
\widehat{\phi _{a}}\right) \}
\end{array}
\end{equation}
where $\Gamma =\Gamma _{L}^{+}+\Gamma _{T}^{+}$ is the partial width for $%
t\rightarrow W^{+}b$. \ The two off-diagonal elements are (read
``upper"/``lower" lines)
\begin{equation}
\widehat{\mathbf{R}}_{\pm \mp }=e^{\pm i\phi _{1}^{t}}\widehat{\mathbf{r}}%
_{\pm \mp }
\end{equation}
with the complex-valued
\begin{equation}
\begin{array}{c}
\widehat{\mathbf{r}}_{+-}=\frac{\Gamma }{4}\{\left[ \left|
R_{+}\right| ^{2}+\left| R_{-}\right| ^{2}\right] \zeta \sin
\theta _{1}^{t}-\cos
\widehat{\theta _{a}}\left[ \left| R_{+}\right| ^{2}-\left| R_{-}\right| ^{2}%
\right] \sigma \sin \theta _{1}^{t} \\ +2\sin \widehat{\theta
_{a}}\left[ \left| R_{+}\right| ^{2}-\left| R_{-}\right|
^{2}\right] (\cos \theta _{1}^{t}[\kappa _{o}\cos \widehat{\phi
_{a}}-\kappa _{o}^{^{\prime }}\sin \widehat{\phi _{a}}]-i[\kappa
_{o}\sin \widehat{\phi _{a}}+\kappa _{o}^{^{\prime }}\cos
\widehat{\phi _{a}}])\}
\end{array}
\end{equation}
and $(\widehat{\mathbf{r}_{-+}})^{\ast
}=\widehat{\mathbf{r}_{+-}}$ . Throughout this paper the symbol $i
\equiv \sqrt {-1}$. Thus we have \ber \widehat{\mathbf{R}}= \left(
\begin{array}{cc}
\widehat{\mathbf{R}}_{++} & e^{i\phi
_{1}^{t}}\widehat{\mathbf{r}_{+-}} \\ e^{-i\phi
_{1}^{t}}\widehat{\mathbf{r}_{-+}} & \widehat{\mathbf{R}}_{--}
\end{array}
\right) \eer

{\bf $CP$-conjugate decay sequence: Only $\bar{b}$-quark
polarimetry:}

For the $CP$-conjugate decay sequence $\bar{t}\rightarrow
W^{-}\bar{b}\rightarrow W^{-}(l^{+}\nu\bar{c})$, we obtain \ber
\overline{\widehat{\mathbf{R}}}=  \left(
\begin{array}{cc}
\overline{\widehat{\mathbf{R}}_{++}} & e^{i\phi _{2}^{t}}\overline{\widehat{%
\mathbf{r}_{+-}}} \\
e^{-i\phi _{2}^{t}}\overline{\widehat{\mathbf{r}_{-+}}} & \overline{\widehat{%
\mathbf{R}}_{--}}
\end{array}
\right) \eer
where $(\overline{\widehat{\mathbf{r}_{-+}}})^{\ast }=\overline{\widehat{%
\mathbf{r}_{+-}}}$ and
\begin{equation}
\begin{array}{c}
\overline{\widehat{\mathbf{R}}_{\pm \pm }}=\frac{\overline{\Gamma }}{4}\{%
\left[ \left| \overline{R_{+}}\right| ^{2}+\left|
\overline{R_{-}}\right| ^{2}\right] \left( 1\mp \overline{\zeta
}\cos \theta _{2}^{t}\right) -\cos \widehat{\theta _{b}}\left[
\left| \overline{R_{+}}\right| ^{2}-\left|
\overline{R_{-}}\right| ^{2}\right] \left( -\overline{\xi }\pm \overline{%
\sigma }\cos \theta _{2}^{t}\right)  \\
\mp 2\sin \widehat{\theta _{b}}\sin \theta _{2}^{t}\left[ \left| \overline{%
R_{+}}\right| ^{2}-\left| \overline{R_{-}}\right| ^{2}\right]
\left( \overline{\kappa _{o}}\cos \widehat{\phi
_{b}}+\overline{\kappa _{o}^{^{\prime }}}\sin \widehat{\phi
_{b}}\right) \}
\end{array}
\end{equation}
\begin{equation}
\begin{array}{c}
\overline{\widehat{\mathbf{r}}_{+-}}=\frac{\overline{\Gamma
}}{4}\{-\left[
\left| \overline{R_{+}}\right| ^{2}+\left| \overline{R_{-}}\right| ^{2}%
\right] \overline{\zeta }\sin \theta _{2}^{t}-\cos \widehat{\theta _{b}}%
\left[ \left| \overline{R_{+}}\right| ^{2}-\left|
\overline{R_{-}}\right| ^{2}\right] \overline{\sigma }\sin \theta
_{2}^{t} \\ +2\sin \widehat{\theta _{b}}\left[ \left|
\overline{R_{+}}\right|
^{2}-\left| \overline{R_{-}}\right| ^{2}\right] (\cos \theta _{2}^{t}[%
\overline{\kappa _{o}}\cos \widehat{\phi _{b}}+\overline{\kappa
_{o}^{^{\prime }}}\sin \widehat{\phi _{b}}]-i[\overline{\kappa
_{o}}\sin \widehat{\phi _{b}}-\overline{\kappa _{o}^{^{\prime
}}}\cos \widehat{\phi _{b}}])\}
\end{array}
\end{equation}

Here we are using the above mentioned ``bar'' notation for the
$CP$
conjugate quantities such as the partial width $\overline{\Gamma }=\overline{%
\Gamma }_{L}^{+}+\overline{\Gamma }_{T}^{+}$ for
$\overline{t}\rightarrow W^{-}\overline{b}$. \ The fundamental
$CP$ relation is
\begin{equation}
B(\lambda _{W^{-}},\lambda _{\overline{b}})=A(-\lambda
_{W^{+}},-\lambda _{b})
\end{equation}
This relationship is useful for constructing ``substitution
rules'' for transcribing to the $CP$ conjugate quantities.  The
results, e.g. the composite decay-density matrices, do not
themselves assume $CP$ and so the S2SC functions, etc. can be used
to test for whether $CP$ holds or not.

\ The helicity parameters for the $CP$ conjugate mode, $\overline{t}%
\rightarrow W^{-}\overline{b}$ include, c.f. lower part of Fig. 2,
\begin{equation}
\overline{\xi }\equiv \frac{1}{\overline{\Gamma }}(\overline{\Gamma }%
_{L}^{-}+\overline{\Gamma }_{T}^{-}),\qquad \overline{\zeta }\equiv \frac{1}{%
\overline{\Gamma }}(\overline{\Gamma }_{L}^{-}-\overline{\Gamma
}_{T}^{-})
\end{equation}
\begin{equation}
\overline{\Gamma }=\overline{\Gamma }_{L}^{+}+\overline{\Gamma }%
_{T}^{+},\qquad \overline{\sigma }\equiv \frac{1}{\overline{\Gamma }}(%
\overline{\Gamma }_{L}^{+}-\overline{\Gamma }_{T}^{+})
\end{equation}
where
\begin{equation}
\overline{\Gamma }_{L}^{\pm }\equiv \left| B(0,\frac{1}{2})\right|
^{2}\pm \left| B(0,-\frac{1}{2})\right| ^{2},\qquad
\overline{\Gamma }_{T}^{\pm
}\equiv \left| B(1,\frac{1}{2})\right| ^{2}\pm \left| B(-1,-\frac{1}{2}%
)\right| ^{2}
\end{equation}
This means that

$\qquad \qquad  \overline{\xi }=$(Prob
$\overline{b}$ is R-handed)
 $ - $  (Prob $\overline{b}$ is L-handed)

$\qquad \qquad \overline{\sigma }=$(Prob $W^{-}$ is
Longitudinally-polarized)  $ - $  (Prob $W^{-}$ is
Transversely-polarized)

The $W^{-}$ polarimetry interference parameters are given by
\begin{equation}
\overline{\omega }\equiv \frac{\overline{I_{{ R}}^{-}}}{\overline{%
\Gamma }},\qquad \overline{\eta }\equiv \frac{\overline{I_{{ R}%
}^{+}}}{\overline{\Gamma }}
\end{equation}
\begin{equation}
\overline{\omega }^{^{\prime }}\equiv \frac{\overline{I_{{ I}}^{-}%
}}{\overline{\Gamma }},\qquad \overline{\eta }^{^{\prime }}\equiv \frac{%
\overline{I_{{ I}}^{+}}}{\overline{\Gamma }}
\end{equation}
where
\begin{equation}
I_{{ R}}^{\pm }\equiv \left| B(0,\frac{1}{2})\right| \left| B(1,%
\frac{1}{2})\right| \cos \overline{\beta _{b}^{R}}\pm \left| B(0,-\frac{1}{2}%
)\right| \left| B(-1,-\frac{1}{2})\right| \cos \overline{\beta
_{b}^{L}}
\end{equation}
\begin{equation}
I_{{ I}}^{\pm }\equiv \left| B(0,\frac{1}{2})\right| \left| B(1,%
\frac{1}{2})\right| \sin \overline{\beta _{b}^{R}}\pm \left| B(0,-\frac{1}{2}%
)\right| \left| B(-1,-\frac{1}{2})\right| \sin \overline{\beta
_{b}^{L}}
\end{equation}

For $B(\lambda _{W^{-}},\lambda \overline{_{b}})=\left| B(\lambda
_{W^{-}},\lambda \overline{_{b}})\right| \exp i\phi _{\lambda
_{W^{-}}}^{bR/L}$, the relative phases are
\begin{eqnarray}
\overline{\alpha _{o}} &=&\phi _{0}^{bL}-\phi _{0}^{bR}\qquad
\qquad \overline{\alpha _{1}}=\phi _{-1}^{b}-\phi _{1}^{b}
\nonumber
\\ \overline{\beta _{b}^{L}} &=&\phi _{-1}^{b}-\phi
_{0}^{bL}\qquad \qquad \overline{\beta _{b}^{R}}=\phi
_{1}^{b}-\phi _{0}^{bR}  \nonumber \\ \overline{\gamma _{-}}
&=&\phi _{-1}^{b}-\phi _{0}^{bR}\qquad \qquad \overline{\gamma
_{+}}=\phi _{1}^{b}-\phi _{0}^{bL}
\end{eqnarray}
We suppress the ``$R/L$" superscript for $\lambda_b = \pm \frac
{1}{2}$  when it is not needed.

The $\overline{b}$-polarimetry helicity parameters for the $CP$
conjugate decay then are
\begin{equation}
\overline{\epsilon _{-}}\equiv \frac{1}{\overline{\Gamma }}\left| B(-1,-%
\frac{1}{2})\right| \left| B(0,\frac{1}{2})\right| \cos \overline{\gamma _{-}%
},\qquad \overline{\epsilon _{-}}^{^{\prime }}\equiv \frac{1}{\overline{%
\Gamma }}\left| B(-1,-\frac{1}{2})\right| \left|
B(0,\frac{1}{2})\right| \sin \overline{\gamma _{-}}
\end{equation}
\begin{equation}
\overline{\epsilon _{+}}\equiv \frac{1}{\overline{\Gamma }}\left| B(1,\frac{%
1}{2})\right| \left| B(0,-\frac{1}{2})\right| \cos \overline{\gamma _{+}}%
,\qquad \overline{\epsilon _{+}}^{^{\prime }}\equiv \frac{1}{\overline{%
\Gamma }}\left| B(1,\frac{1}{2})\right| \left|
B(0,-\frac{1}{2})\right| \sin \overline{\gamma _{+}}
\end{equation}
\begin{equation}
\overline{\kappa _{o}}\equiv \frac{1}{\overline{\Gamma }}\left| B(0,-\frac{1}{%
2})\right| \left| B(0,\frac{1}{2})\right| \cos \overline{\alpha _{o}}%
,\qquad \overline{\kappa _{o}}^{^{\prime }}\equiv \frac{1}{\overline{\Gamma }%
}\left| B(0,-\frac{1}{2})\right| \left| B(0,\frac{1}{2})\right|
\sin \overline{\alpha _{o}}
\end{equation}
\begin{equation}
\overline{\kappa _{1}}\equiv \frac{1}{\overline{\Gamma }}\left| B(-1,-\frac{1}{%
2})\right| \left| B(1,\frac{1}{2})\right| \cos \overline{\alpha _{1}}%
,\qquad \overline{\kappa _{1}}^{^{\prime }}\equiv \frac{1}{\overline{\Gamma }%
}\left| B(-1,-\frac{1}{2})\right| \left| B(1,\frac{1}{2})\right|
\sin \overline{\alpha _{1}}
\end{equation}
In the S2SC distributions and in the single-sided sequential-decay
distributions, sometimes their linear-combinations   $$
\overline{\delta} \equiv \overline{\epsilon _{+}} +
\overline{\epsilon _{-}}, \quad \overline{\epsilon} \equiv
\overline{\epsilon _{+}} - \overline{\epsilon _{-}}, \quad
\overline{\lambda} \equiv \overline{\kappa _{o}} +
\overline{\kappa _{1}}, \quad \overline{\kappa} \equiv
\overline{\kappa _{o}} - \overline{\kappa _{1}}$$ and the
corresponding primed linear-combinations occur, see (59-67).
Simple $CP$ tests were treated in \cite{nklm1}.

{\bf Case: Both $b$-quark polarimetry and $W$-polarimetry:}

We now consider both branches in the decay sequence $t\rightarrow
W^{+}b$ so $b\rightarrow l^{-}\bar{\nu}c$ and
\newline also $W^{+}\rightarrow l^{+}\nu _{l}$ \ [ or
$W^{+}\rightarrow j\overline{_{d}}j_{u}$ ].

We define the ``full'' or double-sequential-decay density matrix for $%
t\rightarrow W^{+}b\rightarrow (l^{+}\nu _{l})(l^{-}\bar{\nu}c)$
by
\begin{equation}
\begin{array}{c}
\not\!{R}_{\lambda _{1}\lambda _{1}^{^{\prime }}}(\theta
_{1}^{t},\phi _{1}^{t};\widetilde{\theta _{a}},\widetilde{\phi
_{a}};\widehat{\theta _{a}}, \widehat{\phi _{a}})  = \sum_{\lambda
_{b}\lambda _{b}^{^{\prime }}}\sum_{\mu _{1}\mu _{1}^{^{\prime
}}}D_{\lambda _{1}\mu _{1}-\lambda _{b}}^{1/2\ast }(\phi
_{1}^{t},\theta _{1}^{t},0)D_{\lambda _{1}^{^{\prime }}\mu
_{1}^{^{\prime }}-\lambda _{b}^{^{\prime }}}^{1/2}(\phi
_{1}^{t},\theta _{1}^{t},0) \\ A(\mu _{1},\lambda _{b})A^{\ast
}(\mu _{1}^{^{\prime }},\lambda _{b}^{^{\prime }}) \widetilde{\rho
}_{\mu _{1}\mu _{1}^{^{\prime }}}(W^{+} \rightarrow l^{+}\nu_{l}
)\hat{\rho}_{\lambda _{b}\lambda _{b}^{^{\prime }}}(b\rightarrow
l^{-}\bar{\nu}c)
\end{array}
\end{equation}
where the summations are over $\lambda _{b}=\pm
\frac{1}{2},\lambda _{b}^{^{\prime }}=\pm \frac{1}{2}$ and the
$W^{+}$ helicities $\mu _{1},\mu _{1}^{^{\prime }}=\pm 1,0$. For
the ``full'' quantities we use a ``slash'' notation. \ Note that
all possible $A(\mu _{1},\lambda _{b})A^{\ast }(\mu _{1}^{^{\prime
}},\lambda _{b}^{^{\prime }})$ interference terms occur in (30).
$\widetilde{\rho }_{\mu _{1}\mu _{1}^{^{\prime }}}$ is given after
(4) and $\hat{\rho}_{\lambda _{b}\lambda _{b}^{^{\prime }}}$  is
(6).

This gives the ``full master-equation''
\begin{equation}
\begin{array}{c}
\not\!{R}_{\lambda _{1}\lambda _{1}^{^{\prime }}}=e^{i(\lambda
_{1}-\lambda _{1}^{^{\prime }})\phi _{1}^{t}} \sum_{\lambda
_{b}\lambda _{b}^{^{\prime }}}\sum_{\mu _{1}\mu _{1}^{^{\prime
}}}\{d_{\lambda _{1},\mu _{1}-\lambda _{b}}^{\frac{1}{2}}(\theta
_{1}^{t})d_{\lambda _{1}^{^{\prime }},\mu _{1}^{^{\prime
}}-\lambda _{b}^{^{\prime }}}^{\frac{1}{2}}(\theta _{1}^{t})A(\mu
_{1},\lambda _{b})A^{\ast }(\mu _{1}^{^{\prime }},\lambda
_{b}^{^{\prime }})  \\
 e^{i(\mu
_{1}-\mu _{1}^{^{\prime }})\widetilde{\phi _{a}}}d_{\mu _{1}1}^{1}(%
\widetilde{\theta _{a}})d_{\mu _{1}^{^{\prime }}1}^{1}(\widetilde{\theta _{a}%
})
e^{i(\lambda _{b}-\lambda _{b}^{^{\prime }})\widehat{\phi _{a}}%
}\sum_{\Lambda =\pm \frac{1}{2}}d_{\lambda _{b},\Lambda }^{\frac{1}{2}}(%
\widehat{\theta _{a}})d_{\lambda _{b}^{^{\prime }},\Lambda }^{\frac{1}{2}}(%
\widehat{\theta _{a}})\left| R_{\Lambda }^{b}(E_{l})\right| ^{2}\}
\end{array}
\end{equation}
It can be expressed in terms of the previously defined helicity
parameters. \ In matrix form the result is
\begin{equation}
\widehat{\mathbf{\not\!{R}}}=\left(
\begin{array}{cc}
\widehat{\mathbf{\not\!{R}}}_{++} & e^{i\phi _{1}^{t}}\widehat{\mathbf{\not{r}}%
_{+-}} \\
e^{-i\phi _{1}^{t}}\widehat{\mathbf{\not{r}}_{-+}} & \widehat{\mathbf{\not\!{R}%
}}_{--}
\end{array}
\right)
\end{equation}
where $(\widehat{\mathbf{\not{r}}_{-+}})^{\ast }=\widehat{\mathbf{%
\not{r}}_{+-}}$ . \ The elements of this matrix are each
conveniently written as the sum of three contributions: \
\begin{equation}
\widehat{\mathbf{\not\!{R}}}=\widehat{\mathbf{\not\!{R}}}^{W}+\widehat{\mathbf{%
\not\!{R}}}^{c}+\widehat{\mathbf{\not\!{R}}}^{s}
\end{equation}
where the first $\widehat{\mathbf{\not\!{R}}}^{W}$ is proportional
to Eq.(12) in \cite{nklm1} which only makes use of $W$%
-polarimetry information, the second
$\widehat{\mathbf{\not\!{R}}}^{c}$ is proportional to the
$b$-polarimetry ``$\cos \widehat{\theta _{a}}$", and the
third  $\widehat{\mathbf{\not\!{R}}}^{s}$ is proportional to the $b$%
-polarimetry ``$\sin \widehat{\theta _{a}}$".

The three contributions to the diagonal elements are:
\begin{equation}
\widehat{\mathbf{\not\!{R}}}_{\pm \pm }^{W}=\frac{1}{8}[\left|
R_{+}\right| ^{2}+\left| R_{-}\right| ^{2}]\mathbf{R}_{\pm \pm }
\end{equation}
where $\mathbf{R}_{\pm \pm }$ is Eq.(13) in \cite{nklm1}.  The
second term in (33) is
\begin{equation}
\begin{array}{c}
\widehat{\mathbf{\not\!{R}}}_{\pm \pm }^{c} = \frac{1}{8}[\left|
R_{+}\right| ^{2}-\left| R_{-}\right| ^{2}]\cos \widehat{\theta
_{a}}{ (}
 { \mathbf{n}%
_{a}^{(-)} } [1\pm { \mathbf{f}_{a}^{(-)} }\cos \theta _{1}^{t}]\mp \frac{1}{\sqrt{2%
}}\sin \theta _{1}^{t}{ \{}\sin 2\widetilde{\theta _{a}}[\eta \cos
\widetilde{\phi _{a}} \\ +\omega ^{^{\prime }}\sin \widetilde{\phi
_{a}}]-2\sin \widetilde{\theta
_{a}}[\omega \cos \widetilde{\phi _{a}}+\eta ^{^{\prime }}\sin \widetilde{%
\phi _{a}}]{ \}}{ )}  \nonumber
\end{array}
\end{equation}
where
\begin{equation}
\left(
\begin{array}{c}
\mathbf{n}_{a}^{(-)} \\ \mathbf{n}_{a}^{(-)}\mathbf{f}_{a}^{(-)}
\end{array}
\right) =-\sin ^{2}\widetilde{\theta _{a}}\frac{\Gamma _{L}^{\mp }}{\Gamma }%
\mp \frac{1}{4}(3+\cos 2\widetilde{\theta _{a}})\frac{\Gamma _{T}^{\mp }}{%
\Gamma }\pm \cos \widetilde{\theta _{a}}\frac{\Gamma _{T}^{\pm
}}{\Gamma }
\end{equation}
with a superscript-tagging per the $l^{-}$ tag of the decaying $b$%
-quark.  Equivalently,
\begin{eqnarray}
\mathbf{n}_{a}^{(-)} &=&\frac{1}{8}(4[1-\sigma ]\cos \widetilde{\theta _{a}}%
-\xi \lbrack 5-\cos 2\widetilde{\theta _{a}}]+\zeta \lbrack 1+3\cos 2%
\widetilde{\theta _{a}}]) \\
\mathbf{n}_{a}^{(-)}\mathbf{f}_{a}^{(-)} &=&\frac{1}{8}(1+3\cos 2\widetilde{%
\theta _{a}}-\sigma \lbrack 5-\cos 2\widetilde{\theta _{a}}]-4[\xi
-\zeta ]\cos \widetilde{\theta _{a}})
\end{eqnarray}
The third term in (33) is
\begin{equation}
\begin{array}{c}
\widehat{\mathbf{\not\!{R}}}_{\pm \pm }^{s} = \frac{1}{8}[\left|
R_{+}\right| ^{2}-\left| R_{-}\right| ^{2}]\sin \widehat{\theta
_{a}}{ \{}\mp \sin
^{2}\widetilde{\theta _{a}}\sin \theta _{1}^{t}{ [}\cos (2\widetilde{%
\phi _{a}}+\widehat{\phi _{a}})\kappa _{1}+2\cos (\widehat{\phi
_{a}})\kappa _{o}  \\ -\sin (2\widetilde{\phi _{a}}+\widehat{\phi
_{a}})\kappa _{1}^{^{\prime }}-2 \sin (\widehat{\phi _{a}})\kappa
_{o}^{^{\prime }}{ ]} \\
+\sqrt{2}\sin \widetilde{\theta _{a}}{ (}\cos (\widetilde{\phi _{a}}+%
\widehat{\phi _{a}}){ [}\delta (1\pm \cos \widetilde{\theta
_{a}}\cos \theta _{1}^{t})+\epsilon (\cos \widetilde{\theta
_{a}}\pm \cos \theta _{1}^{t})   \\ -\sin (\widetilde{\phi
_{a}}+\widehat{\phi _{a}})[\delta ^{^{\prime }}(1\pm \cos
\widetilde{\theta _{a}}\cos \theta _{1}^{t})+\epsilon
^{^{\prime }}(\cos \widetilde{\theta _{a}}\pm \cos \theta _{1}^{t}){ ]}%
{ )\}}
\end{array}
\end{equation}
where $\delta \equiv \epsilon_+ + \epsilon_- ,\epsilon \equiv
\epsilon_+ - \epsilon_-$ and analogously for $\delta^{'}$ and
$\epsilon^{'}$. Note from Fig. 1 that $\delta$ and $\epsilon$ are
$b$-$W$-interference parameters. They only appear in the third
term, i.e. in $\widehat{\mathbf{\not\!{R}}}_{\pm \pm }^{s}$.  The
$b$-polarimetry helicity parameters $\kappa_{o,1}$ also appear
only in this term. Note above in the case of only $b$-polarimetry
the parameters $\kappa_{o,1}$ do appear, but $\delta$ and
$\epsilon$ do not. This is expected since the latter two helicity
parameters only occur due to $b$-$W$-interference.

\bigskip The three contributions to the off-diagonal elements are:
\begin{equation}
\widehat{\mathbf{\not{r}}}_{+-}^{W}=\frac{1}{8}[\left|
R_{+}\right| ^{2}+\left| R_{-}\right| ^{2}]\mathbf{r}_{+-}
\end{equation}
where $\mathbf{r}_{+-}$ is Eq.(14) in \cite{nklm1}.  The second
contribution is
\begin{equation}
\begin{array}{c}
\widehat{\mathbf{\not{r}}}_{+-}^{c} = \frac{1}{8}[\left|
R_{+}\right|
^{2}-\left| R_{-}\right| ^{2}]\cos \widehat{\theta _{a}}{ \{}( { \mathbf{n%
}_{a}^{(-)} }  { \mathbf{f}_{a}^{(-)} } \sin \theta _{1}^{t} \\
-\sqrt{2}\sin \widetilde{\theta _{a}}{ (}\cos \theta
_{1}^{t}[\omega
\cos \widetilde{\phi _{a}}+\eta ^{^{\prime }}\sin \widetilde{\phi _{a}}%
]+i \lbrack \omega \sin \widetilde{\phi _{a}}-\eta ^{^{\prime
}}\cos \widetilde{\phi _{a}}]{ )}   \\ +\frac{1}{\sqrt{2}}\sin
2\widetilde{\theta _{a}}{ (}\cos \theta
_{1}^{t}[\eta \cos \widetilde{\phi _{a}}+\omega ^{^{\prime }}\sin \widetilde{%
\phi _{a}}]+i \lbrack \eta \sin \widetilde{\phi _{a}}-\omega
^{^{\prime }}\cos \widetilde{\phi _{a}}]{ )\}}  \nonumber
\end{array}
\end{equation}
and the third contribution is
\begin{equation}
\begin{array}{c}
\widehat{\mathbf{\not{r}}}_{+-}^{s} = \frac{1}{8}[\left|
R_{+}\right|
^{2}-\left| R_{-}\right| ^{2}]\sin \widehat{\theta _{a}}{ \{}\sqrt{2}%
\sin \theta _{1}^{t}\sin \widetilde{\theta _{a}}{ (}\cos (\widetilde{%
\phi _{a}}+\widehat{\phi _{a}})\epsilon -\sin (\widetilde{\phi _{a}}+%
\widehat{\phi _{a}})\epsilon ^{^{\prime }}
\\
+\cos \widetilde{\theta _{a}}[%
\cos (\widetilde{\phi _{a}}+\widehat{\phi _{a}})\delta -\sin (\widetilde{%
\phi _{a}}+\widehat{\phi _{a}})\delta ^{^{\prime }}]{ )} \\ +\sin
^{2}\widetilde{\theta _{a}}{ (}\cos \theta _{1}^{t}[2\cos
\widehat{\phi _{a}}\kappa _{o}-2\sin \widehat{\phi _{a}}\kappa
_{o}^{^{\prime }}+\cos (2\widetilde{\phi _{a}}+\widehat{\phi
_{a}})\kappa _{1}-\sin (2\widetilde{\phi _{a}}+\widehat{\phi
_{a}})\kappa _{1}^{^{\prime }}]  \\ +i \lbrack -2\sin
\widehat{\phi _{a}}\kappa _{o}-2\cos \widehat{\phi
_{a}}\kappa _{o}^{^{\prime }}+\sin (2\widetilde{\phi _{a}}+\widehat{\phi _{a}%
})\kappa _{1}+\cos (2\widetilde{\phi _{a}}+\widehat{\phi
_{a}})\kappa _{1}^{^{\prime }}]{ )}{ \}}
\end{array}
\end{equation}
Here also, $\delta, \epsilon, \kappa_{o,1}$ and $\delta^{'},
\epsilon^{'}, \kappa_{o,1}^{^{\prime }}$ only appear in
$\widehat{\mathbf{\not{r}}}_{+-}^{s}$ and not in
$\widehat{\mathbf{\not{r}}}_{+-}^{c}$.

In summary, the $\sin \widehat{\theta _{a}}$ dependence of the
``s" superscript terms in $\widehat{\mathbf{\not\!{R}}}$ of (32)
must be used in order to measure the eight $b$-polarimetry
helicity parameters.

{\bf CP-conjugate process: Both $\bar{b}$-quark polarimetry and
$W$-polarimetry:}

For both branches in the $CP$-conjugate decay sequence
$\overline{t}\rightarrow W^{-}\overline{b}$ so
$\overline{b}\rightarrow l^{+}\nu \overline{c}$ and
\newline
also $W^{-}\rightarrow l^{-}\overline{\nu _{l}}$ \ [ or
$W^{-}\rightarrow j_{d}j_{\overline{u}}$ ], we define the ``full''
sequential-decay density matrix by
\begin{equation}
\begin{array}{c}
 \overline{\not\!{R}_{\lambda _{2}\lambda _{2}^{^{\prime }}}}(\theta
_{2}^{t},\phi _{2}^{t};\widetilde{\theta _{b}},\widetilde{\phi
_{b}};\widehat{\theta _{b}}, \widehat{\phi _{b}}) = \sum_{\lambda
_{ \bar{b}}\lambda _{ \bar{b}}^{^{\prime }}}\sum_{\mu _{2}\mu
_{2}^{^{\prime }}}D_{\lambda _{2}\mu _{2}-\lambda _{
\bar{b}}}^{1/2\ast }(\phi _{2}^{t},\theta _{2}^{t},0)D_{\lambda
_{2}^{^{\prime }}\mu _{2}^{^{\prime }}-\lambda _{
\bar{b}}^{^{\prime }}}^{1/2}(\phi _{2}^{t},\theta _{2}^{t},0) \\
B(\mu _{2},\lambda _{ \bar{b}})B^{\ast }(\mu _{2}^{^{\prime
}},\lambda _{ \bar{b}}^{^{\prime }}) \overline{\widetilde{\rho
}}_{\mu _{2}\mu _{2}^{^{\prime }}}(W^{-} \rightarrow
l^{-}\bar{\nu_{l}} )\overline{\hat{\rho}}_{\lambda _{b}\lambda
_{b}^{^{\prime }}}( \bar{b}\rightarrow l^{+}\nu \bar{c})
\end{array}
\end{equation}
where the summations are over $\lambda \overline{_{b}}=\pm \frac{1}{2}%
,\lambda _{\overline{b}}^{^{\prime }}=\pm \frac{1}{2}$ and the
$W^{-}$ helicities $\mu _{2},\mu _{2}^{^{\prime }}=\pm 1,0$.  \
This gives the ``full master-equation''
\begin{equation}
\begin{array}{c}
\overline{\not{\!}{R}}_{\lambda _{2}\lambda _{2}^{^{\prime
}}}=e^{i(\lambda
_{2}-\lambda _{2}^{^{\prime }})\phi _{2}^{t}}\sum_{\lambda _{\overline{b}%
}\lambda _{\overline{b}}^{^{\prime }}}\sum_{\mu _{2}\mu
_{2}^{^{\prime }}}\{d_{\lambda _{2},\mu _{2}-\lambda
\overline{_{b}}}^{\frac{1}{2}}(\theta
_{2}^{t})d_{\lambda _{2}^{^{\prime }},\mu _{2}^{^{\prime }}-\lambda _{%
\overline{b}}^{^{\prime }}}^{\frac{1}{2}}(\theta _{2}^{t})B(\mu
_{2},\lambda
\overline{_{b}})B^{\ast }(\mu _{2}^{^{\prime }},\lambda _{\overline{b}%
}^{^{\prime }}) \\
e^{i(\mu _{2}-\mu _{2}^{^{\prime }})\widetilde{\phi_ b}}d_{\mu _{2},-1}^{1}(%
\widetilde{\theta _{b}})d_{\mu _{2}^{^{\prime
}},-1}^{1}(\widetilde{\theta
_{b}})e^{i(\lambda _{\overline{b}}-\lambda _{\overline{b}}^{^{\prime }})%
\widehat{\phi _{b}}}\sum_{\overline{\Lambda }=\pm
\frac{1}{2}}d_{\lambda
\overline{_{b}},\overline{\Lambda }}^{\frac{1}{2}}(\widehat{\theta _{b}}%
)d_{\lambda _{\overline{b}}^{^{\prime }},\overline{\Lambda }}^{\frac{1}{2}}(%
\widehat{\theta _{b}})\left| \overline{R_{\overline{\Lambda }}^{\overline{b}}%
}(E\overline{_{l}})\right| ^{2}\}
\end{array}
\end{equation}

\qquad It can be expressed in terms of the previously defined
``barred'' helicity parameters. The result is
\begin{equation}
\overline{\widehat{\mathbf{\not{\!}{R}}}}=\left(
\begin{array}{cc}
\overline{\widehat{\mathbf{\not{\!}{R}}}_{++}} & e^{i\phi _{2}^{t}}\overline{%
\widehat{\mathbf{\not{r}}_{+-}}} \\
e^{-i\phi _{2}^{t}}\overline{\widehat{\mathbf{\not{r}}_{-+}}} & \overline{%
\widehat{\mathbf{\not{\!}{R}}}_{--}}
\end{array}
\right)
\end{equation}
where $(\overline{\widehat{\mathbf{\not{r}}_{-+}}})^{\ast }=\overline{%
\widehat{\mathbf{\not{r}}_{+-}}}$ , and \
\begin{equation}
\overline{\widehat{\mathbf{\not{\!}{R}}}}=\overline{\widehat{\mathbf{\not{\!}%
{R}}}^{W}}+\overline{\widehat{\mathbf{\not{\!}{R}}}^{c}}+\overline{\widehat{%
\mathbf{\not{\!}{R}}}^{s}}
\end{equation}
where  $\overline{\widehat{\mathbf{\not{\!}{R}}}^{W}}$ is
proportional to Eq.(17) in \cite{nklm1} ,
$\overline{\widehat{\mathbf{\not{\!}{R}}}^{c}}$ is proportional to
the $\overline{b}$-polarimetry ``$\cos \widehat{\theta _{b}}$'',
and $\overline{\widehat{\mathbf{\not{\!}{R}}}^{s}}$ is
proportional to the $\overline{b}$-polarimetry ``$\sin \widehat{\theta _{b}}$%
''.

The diagonal elements are:
\begin{equation}
\overline{\widehat{\mathbf{\not{\!}{R}}}_{\pm \pm
}^{W}}=\frac{1}{8}[\left|
\overline{R_{+}}\right| ^{2}+\left| \overline{R_{-}}\right| ^{2}]\overline{%
\mathbf{R}_{\pm \pm }}
\end{equation}
where $\overline{\mathbf{R}_{\pm \pm }}$ is Eq.(18) in
\cite{nklm1},
\begin{equation}
\begin{array}{c}
\overline{\widehat{\mathbf{\not\!{R}}}_{\pm \pm }^{c}}
=\frac{1}{8}[\left|\overline{ R_{+}}\right| ^{2}-\left|
\overline{R_{-}}\right| ^{2}]\cos \widehat{\theta _{b}}{ (} - {
\mathbf{n}_{b}^{(+)}  } [ 1 \mp { \mathbf{f}_{b}^{(+)} } \cos
\theta _{2}^{t}]\mp \frac{1}{\sqrt{2}}\sin \theta _{2}^{t}{
\{}\sin 2\widetilde{\theta _{b}}[ \bar{\eta} \cos \widetilde{\phi
_{b}}
\\ - \bar{\omega} ^{^{\prime }}\sin \widetilde{\phi _{b}}]-2\sin
\widetilde{\theta _{b}}[ \bar{\omega} \cos \widetilde{\phi _{b}}-
\bar{\eta} ^{^{\prime }}\sin \widetilde{\phi _{b}}]{ \}}{ )}
\end{array}
\end{equation}
where
\begin{equation}
\left(
\begin{array}{c}
\mathbf{n}_{b}^{(+)} \\ \mathbf{n}_{b}^{(+)}\mathbf{f}_{b}^{(+)}
\end{array}
\right) =-\sin ^{2}\widetilde{\theta _{b}}\frac{\overline{\Gamma _{L}^{\mp }}%
}{\overline{\Gamma }}\mp \frac{1}{4}(3+\cos 2\widetilde{\theta _{b}})\frac{%
\overline{\Gamma _{T}^{\mp }}}{\overline{\Gamma }}\pm \cos
\widetilde{\theta _{b}}\frac{\overline{\Gamma _{T}^{\pm
}}}{\overline{\Gamma }}
\end{equation}
with a superscript-tagging per the $l^{+}$ tag of the decaying $\overline{b}$%
-quark, or
\begin{eqnarray}
\mathbf{n}_{b}^{(+)} &=&\frac{1}{8}(4[1-\overline{\sigma }]\cos \widetilde{%
\theta _{b}}-\overline{\xi }[5-\cos 2\widetilde{\theta _{b}}]+\overline{%
\zeta }[1+3\cos 2\widetilde{\theta _{b}}]) \\
\mathbf{n}_{b}^{(+)}\mathbf{f}_{b}^{(+)} &=&\frac{1}{8}(1+3\cos 2\widetilde{%
\theta _{b}}-\overline{\sigma }[5-\cos 2\widetilde{\theta _{b}}]-4[\overline{%
\xi }-\overline{\zeta }]\cos \widetilde{\theta _{b}})
\end{eqnarray}
and
\begin{equation}
\begin{array}{c}
 \overline{ \widehat{\mathbf{\not\!{R}}}_{\pm \pm }^{s} }=
\frac{1}{8}[\left|\overline{ R_{+}}\right| ^{2}-\left|
\overline{R_{-}}\right| ^{2}]\sin \widehat{\theta _{b}}{ \{}\mp
\sin ^{2}\widetilde{\theta _{b}}\sin \theta _{2}^{t}{ [}\cos
(2\widetilde{\phi _{b}}+\widehat{\phi _{b}}) \bar{\kappa
_{1}}+2\cos (\widehat{\phi _{b}}) \bar{\kappa _{o}} \\ +\sin
(2\widetilde{\phi _{b}}+\widehat{\phi _{b}}) \bar{\kappa
_{1}}^{^{\prime }}+2 \sin (\widehat{\phi _{b}}) \bar{\kappa
_{o}}^{^{\prime }}{ ]} \\ -\sqrt{2}\sin \widetilde{\theta _{b}}{
(}\cos (\widetilde{\phi _{b}}+\widehat{\phi _{b}}){ [}
\bar{\delta} (1\mp \cos \widetilde{\theta _{b}}\cos \theta
_{2}^{t})+ \bar{\epsilon} (\cos \widetilde{\theta _{b}}\mp \cos
\theta _{2}^{t})
\\ +\sin (\widetilde{\phi _{b}}+\widehat{\phi _{b}})[ \bar{\delta} ^{^{\prime
}}(1\mp \cos \widetilde{\theta _{b}}\cos \theta _{2}^{t})+
\bar{\epsilon} ^{^{\prime }}(\cos \widetilde{\theta _{b}}\mp \cos
\theta _{2}^{t}){ ]}{ )\}} \end{array}
\end{equation}

\bigskip The off-diagonal elements are:
\begin{equation}
\overline{\widehat{\mathbf{\not{r}}}_{+-}^{W}}=\frac{1}{8}[\left| \overline{%
R_{+}}\right| ^{2}+\left| \overline{R_{-}}\right| ^{2}]\overline{\mathbf{r}%
_{+-}}
\end{equation}
where $\overline{\mathbf{r}_{+-}}$ is Eq.(19) in \cite{nklm1},
\begin{equation}
\begin{array}{c}
 \overline{ \widehat{\mathbf{\not{r}}}_{+-}^{c}} = \frac{1}{8}[\left|
\overline{ R_{+}}\right| ^{2}-\left| \overline{ R_{-}}\right|
^{2}]\cos \widehat{\theta _{b}}{ \{}( { \mathbf{n}_{b}^{(+)} } {
\mathbf{f}_{b}^{(+)} }\sin \theta _{2}^{t}
\\ -\sqrt{2}\sin \widetilde{\theta _{b}}{ (}\cos \theta
_{2}^{t}[ \bar{\omega} \cos \widetilde{\phi _{b}}- \bar{\eta}
^{^{\prime }}\sin \widetilde{\phi _{b}}]+i \lbrack \bar{\omega}
\sin \widetilde{\phi _{b}}+ \bar{\eta} ^{^{\prime }}\cos
\widetilde{\phi _{b}}]{ )} \\ +\frac{1}{\sqrt{2}}\sin
2\widetilde{\theta _{b}}{ (}\cos \theta _{2}^{t}[ \bar{\eta} \cos
\widetilde{\phi _{b}}- \bar{\omega} ^{^{\prime }}\sin
\widetilde{\phi _{b}}]+i \lbrack  \bar{\eta} \sin \widetilde{\phi
_{b}}+ \bar{\omega} ^{^{\prime }}\cos \widetilde{\phi _{b}}]{ )\}}
\nonumber
\end{array}
\end{equation}
and
\begin{equation}
\begin{array}{c}
 \overline{\widehat{\mathbf{\not{r}}}_{+-}^{s}} = \frac{1}{8}[\left|
\overline{R_{+}}\right| ^{2}-\left|\overline{ R_{-}}\right|
^{2}]\sin \widehat{\theta _{b}}{ \{}\sqrt{2}\sin \theta
_{2}^{t}\sin \widetilde{\theta _{b}}{ (}\cos (\widetilde{\phi
_{b}}+\widehat{\phi _{b}}) \bar{\epsilon} +\sin (\widetilde{\phi
_{b}}+\widehat{\phi _{b}}) \bar{\epsilon} ^{^{\prime }} \\ +\cos
\widetilde{\theta _{b}}[\cos (\widetilde{\phi _{b}}+\widehat{\phi
_{b}}) \bar{\delta} +\sin (\widetilde{\phi _{b}}+\widehat{\phi
_{b}}) \bar{\delta} ^{^{\prime }}]{ )} \\ +\sin
^{2}\widetilde{\theta _{b}}{ (}\cos \theta _{2}^{t}[2\cos
\widehat{\phi _{b}} \bar{\kappa _{o}}+2\sin \widehat{\phi _{b}}
\bar{\kappa _{o}}^{^{\prime }}+\cos (2\widetilde{\phi
_{b}}+\widehat{\phi _{b}}) \bar{\kappa _{1}}+\sin
(2\widetilde{\phi _{b}}+\widehat{\phi _{b}}) \bar{\kappa
_{1}}^{^{\prime }}] \\ +i \lbrack -2\sin \widehat{\phi _{b}}
\bar{\kappa _{o}}+2\cos \widehat{\phi _{b}} \bar{\kappa
_{o}}^{^{\prime }}+\sin (2\widetilde{\phi _{b}}+\widehat{\phi
_{b}}) \bar{\kappa _{1}}-\cos (2\widetilde{\phi
_{b}}+\widehat{\phi _{b}}) \bar{\kappa _{1}}^{^{\prime }}]{ )}{
\}} \end{array}
\end{equation}

The $\sin \widehat{\theta _{b}}$ dependence of the ``s"
superscript terms in $\overline{\widehat{\mathbf{\not\!{R}}}}$ of
(45) must be used in order to measure the eight
$\bar{b}$-polarimetry helicity parameters.

\section{Stage-Two Spin-Correlation Functions \newline Including $b$-Polarimetry}

We include both branches in the decay sequence $t\rightarrow
W^{+}b$ so  $b\rightarrow l^{-}\bar{\nu}c$ and also
$W^{+}\rightarrow l^{+}\nu _{l}$ \ [ or $W^{+}\rightarrow
j\overline{_{d}}j_{u}$ ].  We also include both branches for the
CP-conjugate sequence.

\subsection{The full S2SC function:}

The complete S2SC function is relatively simple in structure even
though it depends on ``5+4+4" variables [c.f. Eq.(66) of
\cite{nklm1}].  Each of the last 4 variables $ \widetilde{\theta
_a},\widetilde{\phi _a};\widehat{\theta _{a}}, \widehat{\phi
_{a}}$  and $\widetilde{\theta _b},%
\widetilde{\phi _b};\widehat{\theta _{b}}, \widehat{\phi _{b}}$
describe the two second-stage branches:
\begin{equation}
\begin{array}{c}
\not{\!}{I}(\Theta _B,\Phi _B;\phi ;\theta _1^t, \widetilde{\theta
_a},\widetilde{\phi _a};\widehat{\theta _{a}}, \widehat{\phi
_{a}};\theta
_2^t,\widetilde{\theta _b},%
\widetilde{\phi _b};\widehat{\theta _{b}}, \widehat{\phi
_{b}})=\sum_{h_1h_2}\{\rho
_{h_1h_2,h_1h_2}^{prod}\not{\!}{R}_{h_1h_1}%
\overline{\not{\!}{R}_{h_2h_2}} \\ +(\rho
_{++,--}^{prod}\not{r}_{+- }\overline{\not{r}_{+-}}+\rho
_{--,++}^{prod}\not{r}_{-+}\overline{\not{r}_{-+}})\cos \phi \\
+i(\rho _{++,--}^{prod}\not{r}_{+-}\overline{\not{r}_{+-}}-\rho
_{--
,++}^{prod}\not{r}_{-+}\overline{%
\not{r}_{-+}})\sin \phi \}
\end{array}
\end{equation}
The ``slashed" composite density matrix elements have been
discussed above.  The production density matrix elements are given
in \cite{nklm1,jack}.

The production density matrices $\rho^{prod}$ in (56) depend on
the angles $\Theta _B,\Phi _B$ which give
\cite{11nklm1,12nklm1,nklm1} the direction of the incident parton
beam, i.e. the quark's momentum or the gluon's momentum, arising
from the incident
$p$ in the $%
p\bar p$, or $pp$, $\rightarrow t\bar tX$ production process:
\begin{equation}
\begin{array}{c}
q \overline{q},  \quad { or } \quad  gg,  \rightarrow
t\overline{t}\rightarrow (W^{+}b)(W^{-}%
\overline{b})
\end{array}
\end{equation}
The important angle between the $t$ and $\bar t$ decay planes is
$$ \phi =\phi _1^t+\phi _2^t = \widehat{\phi} = \widehat{\phi}
_1^t+ \widehat{ \phi} _2^t $$ The other angles have been discussed
previously in consideration of the earlier figures in this paper.
The $\theta _1^t$ angular dependence can be replaced by the
$W^{+}$ energy in the $(t\bar t)_{cm}$ and similarly $\theta _2^t$
by the $W^{-}$ energy \cite{nklm1}.

In the $(t\bar t)_{cm}$ system where $\theta_t$ is the angle
between the $t$-quark momentum and the incident parton-beam this
simplifies to
\begin{equation}
\begin{array}{c}
\not{\!}{I}( \theta _t; \theta _1^t;\widetilde{\theta
_a},\widetilde{\phi _a}; \widehat{\theta _{a}}, \widehat{\phi
_{a}};\theta _2^t; \widetilde{\theta _b}, \widetilde{\phi _b};
\widehat{\theta _{b}}, \widehat{\phi _{b}} )=\sum_{h_1h_2} \rho
_{h_1h_2,h_1h_2}^{prod} (\theta_t) \not{\!}{R}_{h_1h_1}%
\overline{\not{\!}{R}_{h_2h_2}}
\end{array}
\end{equation}
which depends on only the diagonal elements of the ``full"
sequential-decay density matrices $\widehat{\mathbf{R}}_{\pm \pm
}$ and $\overline{\widehat{\mathbf{R}}_{\pm \pm }}$ given
respectively in (32) and (45).  Note that
$\widehat{\mathbf{R}}_{\pm \pm }$   depends on (i) all 8 of the
$W$-polarimetry helicity parameters [ the partial-width $\Gamma$,
the probablility that the emitted $W$ is longitudinally polarized
$P(W_L)=\frac{1}{2} (1+\sigma)$, the probability that the emitted
$b$-quark is L-handed $P(b_L)=\frac{1}{2} (1+\xi)$, $ \zeta$,
$\eta_{L,R},$ and $ {\eta_{L,R}}^{'}$ ] see \cite{nklm1}, and it
also depends on (ii) all 8 of the new $b$-polarimetry parameters [
$\epsilon_{\pm}, \kappa_{o,1}$ and ${\epsilon_{\pm}}^{'}$, $
{\kappa_{o,1}}^{'}$ ].

{\bf Remark:  Use of Alternative-Angles: }The
alternative-angle-labeling of the final $l^{\mp}$ as shown in Fig.
6 can be a significant issue in some circumstances, see
\cite{nklm1,11nklm1} and references therein.  These angles occur
when the boosts to the $b$ and $\bar{b}$ rest frames are directly
from the $(t{\bar{t})_{cm}}$ frame. Recall that this same choice
arises for the labeling of the $W^{\pm}$ decays, see Fig. 4 in
\cite{nklm1}. At a hadron collider, this
alternative-angle-labeling would be useful when both $W^{\pm}$
decay into hadrons. The necessary Wigner-rotation for Fig. 6 is
exactly analogous to that given in \cite{nklm1} in (74,75) with
respect to Fig.4 therein. For the $t$-decay-side, the explicit
``transformation-equations" (involving the Wigner-rotations) are a
simple relabeling of Eqs. (3.22a,b,c) in \cite{11nklm1}, see Fig.
1 therein, and for the $\bar{t}$-decay-side there is an exactly
analogous relabeling of the transformation-Eqs.(3.31a,b,c). At a
future linear collider, use of these alternative-angles for
specifying the final stage-two momenta for some or all of the $b,
\bar{b},W^{\pm}$ decays might also be preferable when $W^+$ and/or
$W^-$ decay leptonically.

In the derivation of S2SC distributions and of single-sided
distributions, some care is needed as to at what step to use the
explicit ``transformation-equations" to the alternative-angles,
see \cite{11nklm1}. Second, the sensitivity in regard to the
measurement of a specific helicity parameter can vary
significantly depending on which minimum variable choice is made.
For instance at an $e^+ e^-$ collider, integrations over some of
the $\theta_{1,2}, \phi_{1,2}$-type variables (which follow after
using the ``transformation-equations") can yield
minimum-variable-distributions which are significantly more
sensitive to some parameters than are the analogous
same-number-of-variable distributions in which the integrations
are performed on the analogous $\theta_{a,b}, \phi_{a,b}$-type
variables, see \cite{nklm1,11nklm1}.

\subsection{A simple two-sided $b$-$W$ spin-correlation function:}

After integrating out the polar angles describing the second-stage
branches, $\widetilde{\theta _a}, \widehat{\theta _{a}};
\widetilde{\theta _b}, \widehat{\theta _{b}} $, we obtain
\begin{equation}
\begin{array}{c}
\not{\!}{I}(\theta _t;\theta _1^t;\widetilde{\phi _a};
\widehat{\phi _{a}};\theta _2^t; \widetilde{\phi _b};
\widehat{\phi _{b}})=\sum_{{q_i}'s, {g_i}'s}\{\rho
_{+-}^{prod}(\theta _t)[\hat{\rho} _{++} \overline{ \hat{\rho}
_{--} }+ \hat{\rho} _{--} \overline{ \hat{\rho} _{++} } ]
\\
+ \rho _{++}^{prod}(\theta _t)[\hat{\rho} _{++} \overline{
\hat{\rho} _{++} }+ \hat{\rho} _{--} \overline{ \hat{\rho} _{--} }
] \}
\end{array}
\end{equation}
The production density matrix elements are given in \cite{nklm1}.

The integrated composite decay density matrix elements are found
to be:   For the $t\rightarrow W^{+}b\rightarrow (l^{+}
\nu_{l})(l^{-}\bar{\nu}c)$ decay sequence, we obtain
$\widehat{\mathbf{\rho }}_{\pm \pm }(\theta
_{1}^{t},\widetilde{\phi _{a}},\widehat{\phi _{a}})$ which depends
on the $t$-quark polar angle and the two second-stage azimuthal
angles
\begin{equation}
\begin{array}{c}
\widehat{\mathbf{\rho }}_{\pm \pm } =\frac{1}{8}[\left|
R_{+}\right| ^{2}+\left| R_{-}\right| ^{2}]{ \{}\frac{2}{3}[1\pm
\zeta \cos \theta
_{1}^{t}]\mp \frac{\pi }{2\sqrt{2}}\sin \theta _{1}^{t}[\eta \cos \widetilde{%
\phi _{a}}+\omega ^{^{\prime }}\sin \widetilde{\phi _{a}}]{ \}}
\\
+\frac{\pi }{32}[\left| R_{+}\right| ^{2}-\left| R_{-}\right| ^{2}]{ %
\{}\frac{\pi }{2\sqrt{2}}{ (}\cos (\widetilde{\phi
_{a}}+\widehat{\phi
_{a}}){ [}\delta \pm \epsilon \cos \theta _{1}^{t}]-\sin (\widetilde{\phi _{a}}+%
\widehat{\phi _{a}})[\delta ^{^{\prime }}\pm \epsilon ^{^{\prime
}}\cos \theta _{1}^{t}]{ )} \\
\mp \frac{2}{3}\sin \theta _{1}^{t}{ [}\cos (2\widetilde{\phi _{a}}+%
\widehat{\phi _{a}})\kappa _{1}+2\cos (\widehat{\phi _{a}})\kappa
_{o}-\sin
(2\widetilde{\phi _{a}}+\widehat{\phi _{a}})\kappa _{1}^{^{\prime }}- 2 \sin (%
\widehat{\phi _{a}})\kappa _{o}^{^{\prime }}{ ]}{ \}}
\end{array}
\end{equation}
For the $\bar{t}\rightarrow W^{-}\bar{b}\rightarrow (l^{-}
\bar{\nu_{l}})(l^{+}\nu\bar{c})$ decay sequence, we obtain $\overline{\widehat{\mathbf{\rho }}_{\pm \pm }}%
(\theta _{2}^{t},\widetilde{\phi _{b}},\widehat{\phi _{b}})$ which
depends on the $\overline{t}$-quark polar angle and the two
second-stage azimuthal angles
\begin{equation}
\begin{array}{c}
\overline{\widehat{\mathbf{\rho }}_{\pm \pm }}%
 =\frac{1}{8}[\left|
\overline{R_{+}}\right| ^{2}+\left| \overline{R_{-}}\right| ^{2}]{ \{}%
\frac{2}{3}[1\mp \overline{\zeta }\cos \theta _{2}^{t}]\pm \frac{\pi }{2%
\sqrt{2}}\sin \theta _{2}^{t}[\overline{\eta }\cos \widetilde{\phi _{b}}-%
\overline{\omega }^{^{\prime }}\sin \widetilde{\phi _{b}}]{ \}}
\\
+\frac{\pi }{32}[\left| \overline{R_{+}}\right| ^{2}-\left| \overline{R_{-}%
}\right| ^{2}]{ \{-}\frac{\pi }{2\sqrt{2}}{ (}\cos (\widetilde{%
\phi _{b}}+\widehat{\phi _{b}}){ [} \overline{\delta} \mp
\overline{\epsilon}\cos \theta
_{2}^{t}]+\sin (\widetilde{\phi _{b}}+\widehat{\phi _{b}})[\overline{\delta }%
^{^{\prime }}\mp \overline{\epsilon }^{^{\prime }}\cos \theta _{2}^{t}]%
{ )} \\
\mp \frac{2}{3}\sin \theta _{2}^{t}{ [}\cos (2\widetilde{\phi _{b}}+%
\widehat{\phi _{b}})\overline{\kappa }_{1}+2\cos (\widehat{\phi _{b}})%
\overline{\kappa }_{o}+\sin (2\widetilde{\phi _{b}}+\widehat{\phi _{b}})%
\overline{\kappa }_{1}^{^{\prime }}+ 2 \sin (\widehat{\phi _{b}})\overline{%
\kappa }_{o}^{^{\prime }}{ ]}{ \}}
\end{array}
\end{equation}

It is important to note that the helicity parameters discussed in
the following section of this paper appear in the above two
density matrices.
The density matrix $\widehat{\mathbf{\rho }}_{\pm \pm }(\theta _{1}^{t},%
\widetilde{\phi _{a}},\widehat{\phi _{a}})$ depends on (i) 3 of
the helicity parameters measurable \cite{nklm1} with only $W$
-polarimetry: $\zeta ,\eta $ and the $\tilde{T}_{FS}$-violating
$\omega ^{^{\prime }}$ parameter which is zero in the SM, and on
(ii) 8 $b$-polarimetry helicity parameters: $\delta \equiv
\epsilon_+ + \epsilon_- ,\epsilon \equiv \epsilon_+ - \epsilon_-
,\kappa _{o},\kappa _{1}$ and the corresponding  primed quantities
which are non-zero if there is $\tilde{T}_{FS}$-violation in
$t\rightarrow W^{+}b$ .

\subsection{Simple single-sided $b$-$W$ distributions:}

Three simple single-sided distributions for $t\rightarrow W^{+}b
\rightarrow ( l^{+}\nu _{l} )( l^{-}\bar{\nu}c) $  ( or
$W^{+}\rightarrow j\overline{_{d}}j_{u}$ ), are the following:
\begin{equation}
\not{\!}{I}(\theta _{t};\widetilde{\theta _{a}},\widehat{\theta _{a}},%
\widetilde{\phi _{a}}+\widehat{\phi _{a}})=\sum_{q_{i},g_{i}}(\rho
_{++}^{prod}+\rho _{--}^{prod})\{\widehat{\not{\!}{R}}_{++}+\widehat{\not{\!}{R}}%
_{--}\}
\end{equation}
where
\begin{equation}
\begin{array}{c}
\{ \widehat{\not{\!}{R}}_{++} + \widehat{\not{\!}{R}}_{--} \}
=\frac{1}{4}[\left| R_{+}\right| ^{2}+\left| R_{-}\right| ^{2}] {
\mathbf{n}_{a}  } +\frac{1}{4}[\left| R_{+}\right| ^{2}-\left|
R_{-}\right| ^{2}]{ \{}\cos \widehat{\theta _{a}} {
\mathbf{n}_{a}^{(-)} }  \\
+\frac{1}{\sqrt{2}}\sin \widehat{\theta _{a}}{ (}\cos (\widetilde{%
\phi _{a}}+\widehat{\phi _{a}})[2\delta \sin \widetilde{\theta _{a}}%
+\epsilon \sin 2\widetilde{\theta _{a}}]  \\ -\sin
(\widetilde{\phi _{a}}+\widehat{\phi _{a}})[2\delta ^{^{\prime
}}\sin \widetilde{\theta _{a}}+\epsilon ^{^{\prime }}\sin 2\widetilde{%
\theta _{a}}]{ )}{ \}}
\end{array}
\end{equation}
Note that this distribution depends on (i) the $W$-polarimetry parameters $%
\sigma ,\xi ,\zeta $ of \cite{nklm1} and on (ii) the
$b$-polarimetry parameters $\delta \equiv \epsilon_+ + \epsilon_-
,\epsilon \equiv \epsilon_+ - \epsilon_- $ and on $\delta
^{^{\prime }},\epsilon ^{^{\prime }}$, see Fig. 1.

By integrating out `` $\cos \widehat{\theta _{a}}"$, we obtain
\begin{equation}
\begin{array}{c}
\not{\!}{I}(\theta _{t};\widetilde{\theta _{a}},\widetilde{\phi _{a}}+%
\widehat{\phi _{a}}) = \frac{1}{2}\int d(\cos \widehat{\theta _{a}}%
)I(\theta _{t};\widetilde{\theta _{a}},\widehat{\theta _{a}},%
\widetilde{\phi _{a}}+\widehat{\phi _{a}})   \\
=\sum_{q_{i},g_{i}}(\rho _{++}^{prod}+\rho _{--}^{prod})
\\
\{ \frac{1}{4}[\left| R_{+}\right| ^{2}+\left| R_{-}\right| ^{2}] { \mathbf{n}_{a} } +%
\frac{\pi }{8\sqrt{2}}[\left| R_{+}\right| ^{2}-\left| R_{-}\right| ^{2}%
]\sin \widetilde{\theta _{a}}{ (}\cos (\widetilde{\phi _{a}}+\widehat{%
\phi _{a}})[\delta +\epsilon \cos \widetilde{\theta _{a}}] \\
-\sin (\widetilde{\phi _{a}}+\widehat{\phi _{a}})[\delta
^{^{\prime }}+\epsilon ^{^{\prime }}\cos \widetilde{\theta _{a}}]{
)} \}
\end{array}
\end{equation}
which displays the difference between $\epsilon ,\epsilon
^{^{\prime }} $ and $\delta ,\delta ^{^{\prime }}$ in the $\cos
\widetilde{\theta _{a}}$ dependence. Next, by also integrating out
`` $\cos \widetilde{\theta _{a}}"$,
\begin{equation}
\begin{array}{c}
\not{\!}{I}(\theta _{t};\widetilde{\phi _{a}}+\widehat{\phi _{a}}) =%
\frac{1}{2}\int d(\cos \widetilde{\theta _{a}})\frac{1}{2}\int
d(\cos
\widehat{\theta _{a}})I(\theta _{t};\widetilde{\theta _{a}},%
\widehat{\theta _{a}},\widetilde{\phi _{a}}+\widehat{\phi _{a}})
\\ =\sum_{q_{i},g_{i}}(\rho _{++}^{prod}+\rho _{--}^{prod})
\\
\{ \frac{1}{6}[\left| R_{+}\right| ^{2}+\left| R_{-}\right| ^{2}] { \mathbf{n}_{a} } +%
\frac{\pi ^{2}}{32\sqrt{2}}[\left| R_{+}\right| ^{2}-\left| R_{-}\right| ^{2}%
]{ (}\delta \cos (\widetilde{\phi _{a}}+\widehat{\phi _{a}})
\\
-\delta ^{^{\prime }}\sin (\widetilde{\phi _{a}}+\widehat{\phi _{a}}){ %
)} \}
\end{array}
\end{equation}
which only depends on $\delta$ and $\delta^{'}$.

For the $CP$-conjugate sequential-decay $\bar{t}\rightarrow W^{-}
\bar{b} \rightarrow ( l^{-}\bar{\nu} _{l} )( l^{+}\nu \bar{c}) $ ,
or $W^{-}\rightarrow j_{d}j_{\bar{u}}$, the distribution analogous
to (62) is:
\begin{equation}
\overline{\not{\!}{I}}(\theta _{t};\widetilde{\theta _{b}},\widehat{\theta _{b}},%
\widetilde{\phi _{b}}+\widehat{\phi _{b}})=\sum_{q_{i},g_{i}}(\rho
_{++}^{prod}+\rho _{--}^{prod})\{ \overline{ \widehat{\not{\!}{R}}}_{++}+ \overline{\widehat{\not{\!}{R}}}%
_{--}\}
\end{equation}
where
\begin{equation}
\begin{array}{c}
\{ \overline{ \widehat{\not{\!}{R}}}_{++} +
\overline{\widehat{\not{\!}{R}}}_{--} \} =\frac{1}{4}[\left|
\overline{R}_{+}\right| ^{2}+\left| \overline{R}_{-}\right| ^{2}]
{ \mathbf{n}_{b}  } +\frac{1}{4}[\left| \overline{R}_{+}\right|
^{2}-\left| \overline{R}_{-}\right| ^{2}]{ \{}\cos \widehat{\theta
_{b}} { \mathbf{n}_{b}^{(+)} }  \\
+\frac{1}{\sqrt{2}}\sin \widehat{\theta _{b}}{ (}\cos (\widetilde{%
\phi _{b}}+\widehat{\phi _{b}})[2\bar{\delta} \sin \widetilde{\theta _{b}}%
+ \bar{\epsilon} \sin 2\widetilde{\theta _{b}}]  \\ -\sin
(\widetilde{\phi _{b}}+\widehat{\phi _{b}})[2\bar{\delta}
^{^{\prime
}}\sin \widetilde{\theta _{b}}+\bar{\epsilon} ^{^{\prime }}\sin 2\widetilde{%
\theta _{b}}]{ )}{ \}}
\end{array}
\end{equation}
where ${ \mathbf{n}_{b}  }$ is Eq.(20-21) in \cite{nklm1}, $ {
\mathbf{n}_{b}^{(+)} }$ is (49-51) above, and $\bar{\delta} \equiv
\bar{\epsilon}_+ + \bar{\epsilon}_- ,\bar{\epsilon} \equiv
\bar{\epsilon}_+ - \bar{\epsilon}_- $.

{\bf Remark:  Use of Single-Sided $b$-$W$ Sequential-Decay
Distributions:} In the context of $b$-polarimetry and of joint
$b$-$W$-polarimetry, especially at a linear collider, it is
important to note what information can be garnered from simpler
single-sided sequential-decay distributions versus S2SC
distributions:  Both the ``full" $11$-angle-variable S2SC
distribution (58)  and the $7$-angle-variable S2SC distribution
(59) depend on all four $b$-polarimetry interference parameters,
$\epsilon_{\pm}, \kappa_{o,1}$ and on the analogous primed
parameters.  The $4$ and $3$-angle-variable single-sided
distributions, (62) and (64), both depend on all of
$\epsilon_{\pm},{\epsilon_{\pm}}^{'}$ but not on any of
$\kappa_{o,1},{\kappa_{o,1}}^{'}$. However, the $2$-angle-variable
single-sided distribution (65) only depends on the sum $\delta=
\epsilon_{+}  +  \epsilon_{-} $ and on $ \delta^{'} $.

These differences can be considered versus single-additional
Lorentz structures where the presence or absence of in signatures
occur similarly for $ \epsilon_{-} $ and $ \kappa_{o} $ (and for
${ \epsilon_{-} }^{'}$ and $ {\kappa_o}^{'} $), and for
$\epsilon_{+}$ and $\kappa_{1}$ (and for $ {\epsilon_{+}}^{'}$ and
${\kappa_{1}}^{'}$). With respect to measurement of only $\delta$
and/or $ {\delta}^{'} $ by the simplest $2$-angle-variable
distribution (65), there is almost complete cancellations in
$\delta$ and ${\delta}^{'}$ in the case of single-additional $V+A,
V$ or $A$ couplings, and some cancellation in ${\delta}^{'}$ for
$f_M - f_E, f_M$ or $f_E$ couplings.  This means that the
$2$-angle-variable single-sided distribution (65) should not be
solely used because the other linear-combinations $ \epsilon =
\epsilon_{+} - \epsilon_{-} $ and ${\epsilon}^{'}$ need also to be
measured. Also, if $ \epsilon $ and $ {\epsilon}^{'} $ are not
measured, then the simple `` $S-P, S$, or $P$ coupling's
signature" of a presence/absence of effects in ($ \epsilon_{-} $
and $ \kappa_{o} $)/($ \epsilon_{+} $ and $ \kappa_{1} $) and
likewise for the primed quantities would not be available.

Obviously, a sensitivity analysis of the ``ideal statistical
errors" and of the systematic errors in regard to the various
helicity parameters in the case of $b$-polarimetry and of joint
$b$-$W$-polarimetry distributions would be useful in the context
of different $\Lambda_b$ polarimetry methods,  the expected number
of events, the details of specific experiments/detectors, and
available experimental information on the dynamics
occurring/not-occurring in top quark decay.

\section{$b$-Polarimetry Interference Parameters
\newline involving $A(-1,- \frac{1}{2} )$}

{\bf Case: In Standard Model and at Ambiguous Moduli Points}

The two $b$-polarimetry interference parameters $\epsilon_+$ and
$\kappa_{o}$ involving the standard model's largest amplitude,
$A(0,- \frac{1}{2} )$ were considered in \cite{cohen}.  Plots for
these parameters were given therein for the case of a
single-additional, real coupling $g_i$. For the other L-handed
b-quark amplitude, $A(-1,- \frac{1}{2} )$, the two analogous
helicity parameters are
\begin{equation}
\begin{array}{c}
\epsilon_{-} \equiv \frac 1\Gamma |A(-1,-\frac 12)||A(0, \frac
12)|\cos \gamma _{-}
\\
= \frac 1\Gamma Re \left\{ A(-1,-\frac 12) A^* (0, \frac 12)
\right\}
\\
\kappa _1 \equiv \frac 1\Gamma |A(1,\frac 12)||A(-1,-\frac
12)|\cos \alpha _1
\\
 =  \frac 1\Gamma Re \left\{ A(1,\frac 12) A^* (-1,
-\frac 12) \right\}
\end{array}
\end{equation}
where  $ \gamma _{-}=\phi _{-1}^L-\phi _0^R $ and  $ \alpha
_1=\phi _1^R-\phi _{-1}^L$, see Fig. 1.

In the SM, the two ${\cal O}(LR)$ helicity parameters which are
between the amplitudes with the largest moduli product are
$\epsilon_+$ and $\kappa_1$, which respective depend on $A(0,-
\frac{1}{2} ) \sim 338$ and \newline $A(-1,- \frac{1}{2} ) \sim
220$ in $g_L = 1$ units. The R-handed amplitudes are $A(1,
\frac{1}{2} ) \sim - 7.16$ and $A(0, \frac{1}{2} ) \sim -2.33$.
Unfortunately, as shown in Table 1, the tree-level values of the
four $b$-interference parameters are only about $1\%$ in the SM,
and also at the $(S+P)$ and $(f_M+f_E)$ ambiguous moduli points
\cite{cohen}.

{\bf Case: ``(V-A) + Single Additional Lorentz Structure" }

Some single-additional Lorentz structures can produce sizable
effects in these four $b$-interference parameters: In
\cite{cohen}, this was shown to occur in $\epsilon_+$ for
additional non-chiral $V,f_M$ or $A,f_E$ couplings, and for the
chiral combinations $V+A,f_M-f_E$ in Figs. 9, and in $\kappa_o$
for additional $V,S,f_M$ or $A,P,f_E$ couplings, and for
$V+A,S-P,f_M-f_E$ in Figs. 10.

In the present paper, analogous plots are given for the
$b$-interference parameters involving the $A(-1,- \frac{1}{2} )$
amplitude.  In Fig. 9 are plots of the $b$-polarimetry
interference parameter $\epsilon_-$ versus $\eta_L$ for the case
of a single-additional, real coupling. By $W$-polarimetry, the
$\eta_L$ parameter can be measured since
\begin{equation}
\begin{array}{c}
\eta _L = \frac{1}{2} ( \eta + \omega ) \equiv \frac 1\Gamma
|A(-1,-\frac 12)||A(0,- \frac 12)|\cos \beta _L
\end{array}
\end{equation}
where $ \beta_L=\phi _{-1}^L- \phi _0^L$ is the relative phase
difference the two helicity amplitudes in (69). Similarly, in Fig.
10 are plots of the $b$-polarimetry interference parameter
$\kappa_1$ versus $\eta_L$. In both sets of plots, Figs. 9 and 10,
the upper(lower) figures are for the case of an additional chiral
(non-chiral) coupling.

Note that an additional $(V-A)$ coupling only effects the overall
magnitude or phase of the $A(\lambda_{W^+}, \lambda_b)$ amplitudes
and so an additional $(V-A)$ coupling will only effect the overall
normalization of the spin-correlation functions. Note also that
due to their L-handed $b$-quark structure, an additional $f_M +
f_E$ or $S+P$ coupling does not significantly effect any of the
four $\epsilon_{\pm}, \kappa_{o,1}$ interference parameters. For
the same reason, these couplings do not significantly effect the
four $ \tilde{T}_{FS}$-violation ${\epsilon_{\pm}}^{'},
{\kappa_{o,1}}^{'}$ parameters, see Figs. 11 and 12 below and see
Figs. 1 and 2 of \cite{adler}.

An additional $S-P$ coupling effects significantly only $\kappa_o$
and $\epsilon_-$. In the non-chiral case, an additional $S$ or $P$
coupling effects significantly only $\kappa_o$ and $\epsilon_-$.
As in \cite{cohen}, the ``oval shapes" of the curves in Figs. 9
and 10 as the coupling strength varies is due to the non-zero
value of the $m_b$ mass, $m_b = 4.5GeV$.

{\bf Case: Explicit $ \tilde{T}_{FS}$ Violation from a
Single-Additional Lorentz Structure}

By ``explicit $ \tilde{T}_{FS}$ violation", we mean an additional
complex-coupling, $ g_i / 2 \Lambda_i $ or $ g_i $, associated
with a specific single-additional Lorentz structure, $ i = S, P, S
\pm P , \ldots $.  For a single-additional gauge-type coupling $V,
A, $ or $V+A$, there is not a significant signature in $
{\eta_L}^{'} $ due to the  $T$-violation ``masking mechanism"
associated with gauge-type couplings \cite{nklm1}. For example:
for an additional pure imaginary $g_R$ coupling plus a purely real
$g_L$, $ {\eta_L}^{'} \sim m_b/m_t $.  In \cite{adler}, we
considered the effects on the ${\epsilon_+}^{'}$ and $
{\kappa_o}^{'}$ helicity parameters of pure-imaginary couplings.
These two parameters involve the $A(0,- \frac{1}{2} )$ amplitude.
Here we consider the analogous effects on ${\epsilon_-}^{'}$ and $
{\kappa_1}^{'}$ which involve the $A(-1,- \frac{1}{2} )$
amplitude.

However, as in \cite{adler}, there are large indirect effects on
other helicity parameters in the case of explicit $
\tilde{T}_{FS}$-violation due to a single-additional
pure-imaginary coupling. Therefore, while the following plots do
show that sizable $ \tilde{T}_{FS}$-violation signatures can occur
due to pure-imaginary additional couplings, such additional
couplings can usually be more simply excluded by $10\%$ precision
measurement of the probabilities $ P(W_L)$ and $P(b_L)$, and of
the $W$-polarimetry interference parameters $\eta$ and $\omega$.

Figs. 11 and 12 display plots of the $b$-polarimetry interference
parameters ${\epsilon_-}^{'}$ and $ {\kappa_1}^{'}$ versus the
coupling strength for the case of a single-additional,
pure-imaginary coupling.  These helicity parameters are defined by
\begin{equation}
\begin{array}{c}
\epsilon_{-}^{\prime } \equiv \frac 1\Gamma |A(-1,-\frac 12)||A(0,
\frac 12)|\sin \gamma _{-}
\\
\kappa _1^{\prime } \equiv \frac 1\Gamma |A(1,\frac
12)||A(-1,-\frac 12)|\sin \alpha _1
\end{array}
\end{equation}
The ``cosine's" of $ \gamma _{-} $ and $ \alpha _1$ occurred above
in (68).

In Figs. 11 and 12 the upper plots (lower plots) are respectively
for the case of an additional non-gauge (gauge) type coupling. The
SM limits are correspondingly at the ``wings" where $ \vert
\Lambda_i  \vert \rightarrow \infty $ ( ``origin" where $g_i
\rightarrow 0 $ ).  As in the case of a purely real additional
coupling, an additional $S-P,S$ or $P$ coupling effects
significantly only $\kappa_o^{\prime }$ and $\epsilon_-^{\prime
}$. In Figs. 11 and 12, the peaks in the curves usually do
correspond to where $\vert \sin \alpha_1 \vert \sim 1$ and $\vert
\sin \gamma_{-} \vert \sim 1$. The exceptions are: in Fig. 11 for
$\epsilon_{-}^{\prime}$ for $f_M, f_E$ where respectively $ \vert
\sin \gamma_- \vert \sim 0.52, 0.48$ at $\vert \Lambda_i \vert =
70GeV$ and for $V,A$ where $ \vert \sin \gamma_- \vert \sim 0.81$
at $\vert g_i \vert = 0.75$; and in Fig. 12 for
$\kappa_{1}^{\prime}$ for $f_M, f_E$ where respectively $ \vert
\sin \alpha_1 \vert \sim 0.52, 0.48$ at $\vert \Lambda_i \vert =
70GeV$ and for $V,A$ where $ \vert \sin \alpha_1 \vert \sim 0.86$
at $\vert g_i \vert = 0.75$. The drops in the curves for small $
\vert \Lambda_i \vert$'s is due to the vanishing of the ``sine" of
the corresponding relative phase.

\section{Concluding Remarks}

The implications and directions for further development of many of
the results in this paper will have to deduced as data accumulates
from model-independent top quark spin-correlation analyses.
However, even if the standard model predictions are correct to
better than the $10\%$ level, top quark decay will still be
uniquely sensitive to ``new" short-distance physics at
nearby-but-not-yet-explored distance scales because of the absence
of hadronization effects and the large $m_t$ mass.

(1)  From analyses only using $W$-polarimetry, it will important
to ascertain the magnitude of the two R-handed $b$-quark
amplitudes for $t \rightarrow W^+ b$. If these are indeed $30$ to
$100$ times smaller than the L-handed $b$-quark amplitudes as
occurs in the SM and occurs at the ($S+P$) and ($f_M+f_E$)
ambiguous moduli points, then the data should show that $[ P(b_L)
= \frac {1}{2} (1+\sigma) ] \simeq 1$, that $ [ 2P(W_L) - P(b_L) =
\frac {1}{2} ( 1 + 2 \sigma - \xi) ] \simeq \zeta $, and that
$\omega \simeq \eta$. Consequently, the $b$-polarimetry
interference parameters $\epsilon_{\pm}, \kappa_{o,1}$, and their
primed analogues, must be small versus their unitarity limit of
$0.5$. In the SM and for the very interesting $(f_M + f_E)$
dynamical ambiguity $ \frac {\epsilon_{+}}{0.5} \sim 3\% $, $\frac
{\kappa_1}{0.5} \sim 2\%$, $\frac {\kappa_o}{0.5} \sim 1\%$ and
$\frac {\epsilon_{-}}{0.5} \sim 0.6 \%$.  In this case, {\bf both}
a large and clean sample of $t$ and $\bar{t}$ decays {\bf and}
mature $\Lambda_b$ polarimetry methods will be required for a
``complete measurement" of the $A(\lambda_{W^+}, \lambda_b)$ and
the $B(\lambda_{W^-}, \lambda_{\bar{b}})$ amplitudes.

However, as shown by the last 4 figures in this paper and by the
analogous ones in \cite{cohen,adler}, the situation is very
different it there are sizable R-handed $b$-quark amplitudes. Such
amplitudes can occur in the case of single-additional Lorentz
structures.  In this case, the use of $\Lambda_b$ polarimetry in
top quark spin-correlation analyses could be uniquely sensitive
and useful in disentangling new physics at the Tevatron and LHC.

(2) Being able to neglect R-handed amplitudes is also important
with respect to searching for $\tilde{T}_{FS}$-violation
signatures. If R-handed amplitudes are negligible, then from
analyses only using $W$-polarimetry,  the data can show that $0
\neq [ {\omega}^{'} \simeq {\eta}^{'} ]$ or that \newline $ 0 \neq
[ {( {\eta_L}^{'} )^2 } \simeq \frac {1}{4} [P( { b_{L} }) + \zeta
][ P( { b_{L} } ) - \zeta ] - ( {\eta_{L} })^2 ] $ as signatures
for $\tilde{T}_{FS}$-violation.  With respect to $b$-polarimetry
information the situation is as discussed in remark (1) above: If
the R-handed amplitudes are $\sim 1\%$ then the four ``primed"
$\tilde{T}_{FS}$-violation $b$-polarimetry parameters will have
magnitudes at most of $\sim 1\%$, c.f. the lower parts of Figs. 1
and 2. However, if R-handed $b$-quark amplitudes are sizable, then
$\tilde{T}_{FS}$-violation signatures can be much larger as shown
by the ``primed" helicity-parameter figures in this paper and in
\cite{adler}.

(3) When final $\bar{\nu}$ angles-and-energy variables are used in
top quark analyses for $b \rightarrow l^- \bar{\nu} c$ decay in
the $\Lambda_b$ mass region, then $\Lambda_b$ polarimetry
signatures will not be suppressed.  This is shown above by a
simple argument in the paragraph following (6), and see Appendix.
Similarly, $\nu$ angles-and-energy variables will not suppress
signatures from $\bar{b} \rightarrow l^+ \nu \bar{c}$ decay.

(4) At the time of a future linear collider, it will be important
to reconsider what information about top quark decays can be
better obtained from (i) the use of the alternative-angles of
Fig.6, and from (ii) the simpler single-sided $b$-$W$-interference
distributions versus S2SC distributions. These issues are briefly
discussed in the ``remarks" following (58) and (67). The
single-sided distributions considered above (62-67) depend on
$\epsilon_{\pm},{\epsilon_{\pm}}^{'}$ but not on
$\kappa_{o,1},{\kappa_{o,1}}^{'}$.

\begin{center}
{\bf Acknowledgments}
\end{center}

We thank L.J. Adler, Jr. for his assistance in preparation of the
helicity parameter plots. We thank experimentalists and theorists
for discussions regarding $\Lambda_b$ polarimetry and HQET.  This
work was partially supported by U.S. Dept. of Energy Contract No.
DE-FG 02-86ER40291.

\begin{appendix}

\section{Appendix:  Formulas for $\left| R_{\Lambda
}^{b}(E_{l})\right|^2$ type factors \newline when $ b\rightarrow
l^{-}\overline{\nu }c$, and $\overline{b}\rightarrow l^{+}\nu
\overline{c}$ :}

In this appendix are listed explicit formulas for the $\left|
R_{\Lambda }^{b}(E_{l})\right|^2$ and $\left|
\overline{R_{\overline{\Lambda} }^{b}}(E_{\bar{l}})\right|^2$
factors, assuming the SM's pure $V-A$ $(V+A)$ couplings
respectively for $b$ $(\bar{b})$ decay. These formulas follow by
use of [17], the second paper in [8] and its references.

\qquad  For the decay $ b\rightarrow l^{-}\overline{\nu }c$ , in
(6)
\begin{equation}
\left| R_{\pm }^{b}\right| ^{2}=R(x_{l})\mp S(x_{l})
\end{equation}
where
\begin{equation}
R(x_{l})=\frac{1}{f(\varepsilon
_{c})}\frac{x_{l}^{2}(1-\varepsilon
_{c}-x_{l})^{2}}{(1-x_{l})^{2}}\left[ 3-2x_{l}+\varepsilon _{c}\left( \frac{%
3-x_{l}}{1-x_{l}}\right) \right]
\end{equation}
\begin{equation}
S(x_{l})=\frac{1}{f(\varepsilon
_{c})}\frac{x_{l}^{2}(1-\varepsilon
_{c}-x_{l})^{2}}{(1-x_{l})^{2}}\left[ -1+2x_{l}+\varepsilon _{c}\left( \frac{%
1+x_{l}}{1-x_{l}}\right) \right]
\end{equation}
with $x_{l}=2E_{l}/m_{b}$, $\varepsilon _{c}=m_{c}^{2}/m_{b}^{2}$, and  $%
f(\varepsilon _{c})=1-8$ $\varepsilon _{c}+8\varepsilon
_{c}^{2}-\varepsilon
_{c}^{4}-12\varepsilon _{c}^{2}\log \varepsilon _{c}$. \ For the $CP$-conjugate mode $%
\overline{b}\rightarrow l^{+}\nu \overline{c}$ , in (15-16, 48-55)
\begin{equation}
\left| \overline{R_{\pm }^{b}}\right| ^{2}=R(\overline{x_{l}})\pm S(%
\overline{x_{l}})
\end{equation}
with
\begin{eqnarray}
\overline{\hat{\rho}_{\lambda _{\bar{b}}\lambda
_{\bar{b}}^{^{\prime }}}}(\overline{b}
&\rightarrow & l^{+}\nu \overline{c})=\overline{\hat{\rho}_{\lambda _{%
\overline{b}}\lambda _{\overline{b}}^{^{\prime }}}}(\widehat{\phi _{b}},%
\widehat{\theta _{b}},E\overline{_{l}})  \nonumber \\
&=&\sum_{\overline{\Lambda }=\pm \frac{1}{2}}D_{\lambda \overline{_{b}}%
\overline{\Lambda }}^{1/2\ast }(\widehat{\phi _{b}},\widehat{\theta _{b}}%
,0)D_{\lambda _{\overline{b}}^{^{\prime }}\overline{\Lambda }}^{1/2}(%
\widehat{\phi _{b}},\widehat{\theta _{b}},0)\left| \overline{R_{\overline{%
\Lambda }}^{b}(E\overline{_{l}})}\right| ^{2}
\end{eqnarray}

The $\bar{\nu}$ (or $\nu$) angle-energy-spectra is very useful in
$\Lambda _{b}$ polarimetry methods, see [3-6, 8-10] and remarks
above in paragraph after (6).  Only simple changes are needed in
the present formalism to use $\bar{\nu}$ (or $\nu$)
angles-and-energy variables: \ For describing $ b\rightarrow
l^{-}\overline{\nu }c$, the angles $\widehat{\varphi
}_{a}^{^{\prime }}$, $\widehat{\theta } _{a}^{^{\prime }}$ can be
used to label the anti-neutrino momentum direction in place of the
$l^-$ angles $ \widehat{\varphi }_{a}$, $\widehat{\theta }_{a}$.
This is only a matter of adding ``primes'' to these angles in the
various expressions. In place of (6) one has
\begin{eqnarray}
\hat{\rho}_{\lambda _{b}\lambda _{b}^{^{\prime }}}(b &\rightarrow & l^{-}%
\bar{\nu}c)=\hat{\rho}_{\lambda _{b}\lambda _{b}^{^{\prime
}}}(\widehat{\phi }_{a}^{^{\prime }},\widehat{\theta
}_{a}^{^{\prime }},E_{\overline{\nu }}) \nonumber \\
&=&\sum_{\Lambda =\pm \frac{1}{2}}D_{\lambda _{b}\Lambda }^{1/2\ast }(%
\widehat{\phi }_{a}^{^{\prime }},\widehat{\theta }_{a}^{^{\prime
}},0)D_{\lambda _{b}^{^{\prime }}\Lambda }^{1/2}(\widehat{\phi }%
_{a}^{^{\prime }},\widehat{\theta }_{a}^{^{\prime }},0)\left|
N_{\Lambda }^{b}(E_{\overline{\nu }})\right| ^{2}
\end{eqnarray}
where
\begin{equation}
\left| N_{\pm }^{b}\right| ^{2}=U(y_{\nu })\mp V(y_{\nu })
\end{equation}
where $V(y_{\nu })=-U(y_{\nu })$ and
\begin{equation}
U(x_{l})=\frac{1}{f(\varepsilon _{c})}\frac{6y_{\nu
}^{2}(1-\varepsilon _{c}-y_{\nu })^{2}}{(1-y_{\nu })}
\end{equation}
with $y_{\nu }=2E_{\overline{\nu }}/m_{b}$. Therefore, in the
expressions in the text, when the $\bar{\nu}$ variables are used,
$\left| R_{+} \right| ^{2} $ of (9) is replaced by $2 U(y_{\nu})$
and $\left| R_{-}\right| ^{2}$ is set equal to zero. This
vanishing is occurring because in the $b$-quark rest frame,
$\Lambda $ in $\left| N_{\Lambda }^{b}(E_{\overline{\nu }})\right|
^{2}$ is the eigenvalue of $\mathbf{J\cdot }\widehat{\mathbf{p}}_{\overline{%
\nu }} $ where $\mathbf{J}$ is the angular momentum operator.
Therefore, for $m_{\nu }=0$ , the final $\overline{\nu }$ is
purely R-handed so $\left| N_{-}^{b}\right| ^{2}=0$.  See
paragraph after (5) in \cite{jd1}.

Similarly for the neutrino in $\overline{b}\rightarrow l^{+}\nu
\overline{c}$ : \ In place of \ the angles $\widehat{\varphi
}_{b}$, $\widehat{\theta }_{b}
$ , the angles $\widehat{\varphi }_{b}^{^{\prime }}$, $\widehat{\theta }%
_{b}^{^{\prime }}$ can be used to label the neutrino momentum
direction.  Then in place of (75) one has
\begin{eqnarray}
\overline{\hat{\rho}_{\lambda _{\bar{b}}\lambda
_{\bar{b}}^{^{\prime }}}}(\overline{b} &\rightarrow & l^{+}\nu
\overline{c})=\overline{\hat{\rho}_{\lambda
_{\bar{b}}\lambda _{\bar{b}}^{^{\prime }}}}(\widehat{\phi }_{b}^{^{\prime }},\widehat{%
\theta }_{b}^{^{\prime }},E_{\nu })  \nonumber \\
&=&\sum_{\overline{\Lambda }=\pm \frac{1}{2}}D_{\lambda _{\overline{b}}%
\overline{\Lambda }}^{1/2\ast }(\widehat{\phi }_{b}^{^{\prime }},\widehat{%
\theta }_{b}^{^{\prime }},0)D_{\lambda _{\overline{b}}^{^{\prime }}\overline{%
\Lambda }}^{1/2}(\widehat{\phi }_{b}^{^{\prime }},\widehat{\theta }%
_{b}^{^{\prime }},0)\left| \overline{N_{\bar{\Lambda}
}^{b}}(E_{\nu })\right| ^{2}
\end{eqnarray}
where
\begin{equation}
\left| \overline{N_{\pm }^{b}}\right| ^{2}=U(\overline{y_{\nu }})\pm V(%
\overline{y_{\nu }})
\end{equation}
When the $\nu$ variables are used, $\left| \overline{R_{-}}
\right| ^{2} $ is replaced by $2 U(\overline{y_{\nu}})$ and
$\left| \overline{R_{+}}\right| ^{2} = 0$ because the final
$\bar{\nu}$ is purely L-handed.
\end{appendix}

\begin{center}
{\bf Table Captions}
\end{center}

Table 1: For the standard model and at the ($S+P$) and ($f_M+f_E$)
ambiguous moduli points, numerical values of $\eta_L$ and of the
four $b$-polarimetry interference parameters, $\epsilon_{\pm},
\kappa_{o,1}$ which are defined by the lower sketch in Fig. 1.
[$m_t=175GeV, \; m_W = 80.35GeV, \; m_b = 4.5GeV$ ]

\begin{center}
{\bf Figure Captions}
\end{center}

FIG. 1: For $ t\rightarrow W^{+}b $ decay, a display of the four
helicity amplitudes  $A(\lambda _{W^{+}},\lambda _b)=$
\newline $|A|\exp (i\phi _{\lambda _{W^{+}}}^{L/R})$ relative to the
$W^+$ and $b$ helicities. The upper sketch defines the measurable
relative phases and the lower sketch defines their corresponding
real-part and imaginary-part (primed) helicity parameters.
Throughout this paper the symbol $i \equiv \sqrt {-1}$.   For a
pure $V-A$ coupling, the $\beta $'s vanish and all the $\alpha $'s
and $\gamma $'s equal $+\pi $ (or $-\pi $) to give the overall
minus sign in each of the standard model's R-handed $b$-quark
amplitudes, see \cite{cohen,adler}.

FIG. 2: For $ \bar{t} \rightarrow W^{-}\bar{b} $ decay (the
$CP$-conjugate process), the relative phases and associated
helicity parameters are defined as in Fig. 1 but now with
``barred" accents. Compare (18-29). Now the R-handed $\bar{b}$
amplitudes, $B(\lambda_{W^-}, \frac {1}{2})$ reference the
$\bar{\alpha}_0, \bar{\alpha}_1$ relative-phase-directions,
$B(\lambda _{W^{-}},\lambda \overline{_{b}})=\left| B \right| \exp
i\phi _{\lambda _{W^{-}}}^{bR/L}$.

FIG. 3:  The three angles $\theta _1^t$, $\theta _2^t$ and $\phi $
describe the first stage in the sequential-decays of the $(t\bar
t)$ system in which $%
t\rightarrow W^{+}b$ and $\bar t\rightarrow W^{-}\bar b$.

FIG. 4:  For the sequential decay $t\rightarrow W^{+}b$ followed
by
 $b\rightarrow l^{-}\bar{\nu} X $, the two pairs of spherical angles
$\widehat{\theta _1^t}$,
$\widehat{\phi _1^t}$ and $%
\widehat{\theta _a}$,$\widehat{\phi _a}$ describe respectively the
$b$ momentum in the ``first stage" $t\rightarrow W^{+}b$ and the
$l^{-}$ momentum in the ``second stage" $b\rightarrow
l^{-}\bar{\nu} X $. Angles associated with the
 $b$, or ${\Lambda_b}$, branching's momenta directions have ``hat" accents
whereas those associated with the $W^{+}$ branching have ``tilde"
accents, as in Fig. 7 below. The
spherical angles $%
\widehat{\theta _a}$, $\widehat{\phi _a}$ specify the  $l^{-}$
momentum in the $b$ rest frame when the boost is from the $t_1$
rest frame. In this figure, $\widehat{\phi _1^t}$ is shown equal
to zero for simplicity of illustration.  The positive
$\widehat{x_a}$ direction is specified by the $\bar{t}$ momentum
direction.

FIG. 5: This figure is symmetric versus Fig. 4.  The spherical
angles $\widehat{\theta _b} $, $\widehat{\phi _b}$ specify the
$l^{+ }$  momentum in the $\bar{b}$ rest frame when the boost is
from the $\bar t_2$ rest frame.

FIG. 6: The spherical angles $\widehat{\theta _1}$,$\widehat{\phi
_1}$ specify the $l^{-}$ momentum in the $b$ rest frame when the
boost is directly from the $(t\bar t)_{cm}$ frame. Similarly,
$\widehat{\theta _2}$, $\widehat{\phi _2}$ specify the $l^{+}$
momentum in the $\bar{b}$ rest frame. The $b \bar{b}$ production
half-plane specifies the positive $\widehat{x_1}$ and
$\widehat{x_2}$ axes.

FIG. 7:  This figure labels the $W^{+}$ branch analogous to the
labels in Fig. 4 for the $b$ branch. The two pairs of spherical
angles $\theta _1^t$,
$\phi _1^t$ and $%
\widetilde{\theta _a}$,$\widetilde{\phi _a}$ describe the
respective stages in the sequential decay $t\rightarrow W^{+}b$
followed by $W^{+}\rightarrow j_{\bar d}j_u$ [ or
$W^{+}\rightarrow l^{+}\nu $ ].  The spherical angles
$\widetilde{\theta _a} $, $\widetilde{\phi _a}$ specify the
$j_{\bar{d}}$ jet [or the $l^{+ }$] momentum in the $W^{+}$rest
frame.

FIG. 8: The labels are as in Fig. 7 but here for the $W^{-}$
branch. The spherical angles $\widetilde{\theta _b} $,
$\widetilde{\phi _b}$ specify the $j_d$ jet [or the $l^{- }$]
momentum in the $W^{-}$rest frame.

FIG. 9: Plots of the $b$-polarimetry interference parameter
$\epsilon_-$ versus $\eta_L $ for the case of a single-additional,
real coupling. The SM prediction is shown by the solid rectangle.
The {\bf upper plot} is for a single-additional, real chiral
coupling.  The value of $\epsilon_- \sim 0$ for the other
couplings $S+P, f_M + f_E$.   The {\bf lower plot} is for a
single-additional, real non-chiral coupling.  The omitted curves
for $A, P, f_E $ are respectively almost mirror images about the
$\eta_L$ axis. Coupling strengths at representative points are
given in the table associated with the respective plot.   The
unitarity limit is a circle of radius $0.5$, centered at the
origin, in each of these plots and in those in Fig. 10.

FIG. 10: Plots of the $b$-polarimetry interference parameter
$\kappa_1$ versus $\eta_L $ for the case of a single-additional,
real coupling.  The {\bf upper plot} is for a single-additional,
real chiral coupling; the value of $\kappa_1 \sim 0$ for the
omitted couplings $S \pm P, f_M + f_E$.  The {\bf lower plot} is
for a single-additional, real non-chiral coupling; the omitted
curves for $A, f_E $ are respectively almost mirror images about
the $\eta_L$ axis.  The value of $\kappa_1 \sim 0$ for the omitted
couplings $S, P$.

FIG. 11:   Plots of the $b$-polarimetry,
$\tilde{T}_{FS}$-violation, interference parameter
${\epsilon_-}^{'}$ versus coupling strength for the case of a
single-additional {\bf pure-imaginary coupling}.  Curves are for
non-gauge-type couplings ({\bf upper plot}), gauge-type couplings
({\bf lower plot}), versus respectively the effective-mass scale
$\Lambda_i$, or coupling strength $g_i$ in $g_L =1$ units.  Curves
are omitted in these plots and in the following Fig. 12 when the
couplings produce approximately zero deviations in the helicity
parameter of interest.

FIG. 12: Plots of the $b$-polarimetry, $\tilde{T}_{FS}$-violation,
interference parameter ${\kappa_1}^{'}$ versus coupling strength
for the case of a single-additional  {\bf pure-imaginary
coupling}.  Curves are for non-gauge-type coupling ({\bf upper
plot}), gauge-type coupling ({\bf lower plot}), versus
respectively the effective-mass scale $\Lambda_i$, or coupling
strength $g_i$ in $g_L =1$ units.

\end{document}